\documentclass[a4paper,fleqn,usenatbib]{mnras}

\usepackage{natbib}
\usepackage{multicol}

\usepackage{graphicx}	
\usepackage{amsmath}	
\usepackage{amssymb}	
\usepackage{subcaption}
\usepackage{float}
\usepackage{multirow}
\usepackage{bm}
\usepackage{mathabx}

\DeclareGraphicsExtensions{.pdf,.png,.jpg,.eps}
\graphicspath{{figures/}}

\newcommand{\wgg}{\ensuremath{w_{gg}}}

\newcommand{\wgp}{\ensuremath{w_{g+}}}

\newcommand{\xigp}{\ensuremath{\xi_{g+}}}
\newcommand{\xipp}{\ensuremath{\xi_{++}}}

\newcommand{\mpch}{\ensuremath{h^{-1}\text{Mpc}}}

\def\reff@jnl#1{{\rm#1\/}}

\def\aj{\reff@jnl{AJ}}                  
\def\araa{\reff@jnl{ARA\&A}}            
\def\apj{\reff@jnl{ApJ}}                        
\def\apjl{\reff@jnl{ApJ}}               
\def\apjs{\reff@jnl{ApJS}}              
\def\apss{\reff@jnl{Ap\&SS}}            
\def\aap{\reff@jnl{A\&A}}               
\def\aapr{\reff@jnl{A\&A~Rev.}}         
\def\aaps{\reff@jnl{A\&AS}}             
\def\baas{\reff@jnl{BAAS}}              
\def\jrasc{\reff@jnl{JRASC}}            
\def\memras{\reff@jnl{MmRAS}}           
\def\mnras{\reff@jnl{MNRAS}}            
\def\physrep{\reff@jnl{Phys.Rep.}}
\def\pra{\reff@jnl{Phys.Rev.A}}         
\def\prb{\reff@jnl{Phys.Rev.B}}         
\def\prc{\reff@jnl{Phys.Rev.C}}         
\def\prd{\reff@jnl{Phys.Rev.D}}         
\def\prl{\reff@jnl{Phys.Rev.Lett}}      
\def\pasp{\reff@jnl{PASP}}              
\def\pasj{\reff@jnl{PASJ}}              
\def\skytel{\reff@jnl{S\&T}}            
\def\solphys{\reff@jnl{Solar~Phys.}}    
\def\sovast{\reff@jnl{Soviet~Ast.}}     
\def\ssr{\reff@jnl{Space~Sci.Rev.}}     
\def\nat{\reff@jnl{Nature}}             

\newcommand{\beq}{\begin{equation}}
\newcommand{\eeq}{\end{equation}}
\newcommand{\beqa}{\begin{eqnarray}}
\newcommand{\eeqa}{\end{eqnarray}}

\usepackage[usenames,dvipsnames,svgnames,x11names]{xcolor}
\definecolor{sdeep}{rgb}{0., 0.6, 0.}

\usepackage{orcidlink}
\newcommand{\refresponse}[1]{#1}

\title[Multipole IA estimators]{Increasing the power of \refresponse{weak lensing} data with multipole-based intrinsic alignment estimators
}

\author[S.~Singh et~al.]{
Sukhdeep Singh$^{1}$\orcidlink{https://orcid.org/0000-0001-6289-9208},
Ali Shakir$^{1}$, 
Yesukhei Jagvaral$^{1}$\orcidlink{https://orcid.org/0000-0001-7068-7037}, 
Rachel Mandelbaum$^{1}$\thanks{rmandelb@andrew.cmu.edu}\orcidlink{https://orcid.org/0000-0003-2271-1527}\\
\\
   $^{1}$McWilliams Center for Cosmology, Department of Physics, Carnegie Mellon University, Pittsburgh, PA 15213, USA
}

\date{Accepted XXX. Received YYY; in original form ZZZ}

\pubyear{2023}

\begin{document}
\label{firstpage}
\pagerange{\pageref{firstpage}--\pageref{lastpage}}
\maketitle

\begin{abstract}
%
It has long been known that galaxy shapes align coherently with the large-scale density field.  Characterizing this effect is essential to interpreting measurements of weak gravitational lensing, the deflection of light from distant galaxies by matter overdensities along the line of sight, as it \refresponse{also} produces coherent galaxy alignments that we wish to interpret in terms of a cosmological model.  Existing direct measurements of intrinsic alignments using galaxy samples with high-quality shape and redshift measurements typically use well-understood but sub-optimal projected estimators, which do not make good use of the information in the data when comparing those estimators to theoretical models.  We demonstrate a more optimal estimator, based on a multipole expansion of the correlation functions or power spectra, for direct measurements of galaxy intrinsic alignments. We show that even using the lowest order multipole alone increases the significance of inferred model parameters using simulated and real data, without any additional modeling complexity. We apply this estimator to \refresponse{measurements of parameters of the non-linear alignment model using data} from the Sloan Digital Sky survey, demonstrating consistent results with a factor of $\sim$2 greater precision in parameter fits to intrinsic alignments models. This result is functionally equivalent to quadrupling the survey area, but without the attendant costs -- thereby demonstrating the value in using this new estimator in current and future intrinsic alignments measurements using spectroscopic galaxy samples. 

\end{abstract}




\section{Introduction}

Weak gravitational lensing, the deflection of light rays from distant galaxies due to the gravitational potential of intervening matter, is among the most promising ways to stress test and constrain the parameters of our current cosmological model \citep{Weinberg2013,Kilbinger2015}.  In recent years, weak lensing has been used to constrain the amplitude of structure growth and its evolution with a precision of 3--5 per cent \citep{Amon2022,Secco2022,vandenBusch2022,Dalal2023,Li2023}.  Upcoming surveys such as \refresponse{the {\em Euclid} satellite} \citep{Laureijs2011}, the Vera C.\ Rubin Observatory Legacy Survey of Space and Time \citep{Ivezic2019}, and the Nancy Grace Roman Space Telescope High Latitude Imaging Survey \citep{akeson19} have been formulated with a goal of increasingly ambitious weak lensing measurements to enable precise cosmological measurements.

With increasing statistical constraining power, characterizing and mitigating sources of systematic uncertainty in weak lensing \citep{Mandelbaum2018} is increasingly a priority.  Since measuring weak lensing relies on measuring coherent galaxy alignments, any alignments not due to lensing must be either removed or modeled.  One such source of coherent alignments other than lensing are {\em intrinsic alignments} \citep{Joachimi2015ia,Troxel2015}, the coherent alignment of galaxy shapes due to physically localized effects such as tidal fields generated by the large-scale structure.  Since the first detections of large-scale alignments of galaxy shapes with overdensities \citep{Mandelbaum2006,Hirata2007}, the field has expanded to include a wide range of measurement and modeling efforts.  Direct measurement of intrinsic alignments using galaxy samples with spectroscopic or high-quality photometric redshifts, and with galaxy shape estimates, is of particularly high value as it provides a direct empirical constraint on the amplitude and scale-dependence of these alignments, and their dependence on galaxy properties \citep[for recent constraints, see for example][]{Johnston2019, Fortuna2021, Samuroff2022}. 

Two major challenges in direct measurement of alignments is in the quantity and type of data available to carry them out.  Current direct alignment measurements are dominated by the noise from the random component of galaxy shapes, and typically can constrain the alignments over tens of Mpc at the $\sim$10 per cent level.  Procuring larger samples with both robust shape and \refresponse{high-quality redshift measurements} is quite challenging given the cost of the latter, imposing a practical limitation on the size of available datasets and on the increase in precision that may be achieved through larger datasets.  Another challenge relates to the fact that the existing samples are typically brighter than those used for weak lensing measurements, introducing some systematic uncertainty in extrapolating the trends found in direct measurements  to the samples used for weak lensing measurements.

Given limitations in the sizes of suitable datasets, and a desire to split existing samples by properties such as luminosity and color to more deeply understand variations in alignment with these quantities, a natural question is how to make and interpret the measurements in a way that optimally extracts information from them.  Many of those measurements  have used 2D projected correlation functions in order to reduce the burden of modeling redshift-space distortions and other sources of complexity, though a few have used other statistics \citep[e.g.,][]{Okumura2009}. Carrying out the projected correlation function measurement involves measuring density-shape correlations as a function of both projected separation $r_\mathrm{p}$ and line-of-sight separation $\Pi$, then integrating along some range of $\Pi$ values.  However, \citet{Singh2016ia} showed that basic geometric considerations lead to a well-defined scaling of the alignment signal in the $(r_\mathrm{p}, \Pi)$ plane, and this opens up  the possibility of a more optimal matched filter approach with a multipole expansion in that plane \citep[see also][]{Okumura2020}.  Doing so has the potential to not only increase the statistical significance of intrinsic alignment model constraints and trends with galaxy properties, but also to improve our ability to constrain the intrinsic alignment model on intermediate scales where we expect behavior that requires more complex modeling efforts \citep[e.g.,][]{Blazek2015,2020JCAP...01..025V,Fortuna2021hm}. 

The goal of this work, therefore, is to demonstrate an improved intrinsic alignment estimator that increases the statistical precision of intrinsic alignment model constraints.  We quantify the efficacy of this estimator using simulated data along with a re-analysis of data from the Sloan Digital Sky Survey (SDSS). \refresponse{The simulated data allows for controlled tests separating out different physical effects in a way that cannot be done in real data -- for example, comparing configuration and redshift space.  However, use of real data is important to confirm the quantitative improvements in performance of the new estimator in reality.  Since intrinsic alignments depend on the galaxy population, and the galaxy populations are quite different between the simulations and data, we cannot meaningfully compare the measured IA in the simulations and data with each other.  However, our primary goals are to illustrate (a) the statistical gains in IA model constraints with a multipole-based estimator, and (b) the statistical consistency of the best-fitting IA model parameters with the multipole-based estimator versus traditional estimators.  This can be achieved via an internal comparison of the results for a specific SDSS sample using two different estimators.}

\refresponse{The potential gains from  improved estimates of intrinsic alignment parameters are significant.  While the details depend on galaxy samples and other aspects of the survey (area, depth, etc.) and modeling (intrinsic alignments model used), a cosmological forecast from \citet{Johnston2019} showed that replacing the commonly-used uninformative priors on intrinsic alignment parameters with informative priors improved constraints on cosmological parameters by up to 60 per cent.  This motivates improving constraints on intrinsic alignments model parameters through a variety of means, including the improved estimators developed in this paper.}

The outline of this work is as follows.  In Sec.~\ref{sec:formalism}, we present the formalism used to describe intrinsic alignments and to connect our estimators to theory.  Next, we describe the simulated and real datasets used in this work in Sec.~\ref{sec:data}.  Our measurement methods are described in Sec.~\ref{sec:methods}\refresponse{.}  Sec.~\ref{sec:results} shows our results, and we discuss them and their implications for future intrinsic alignments measurements in Sec.~\ref{sec:conclusions}.


\section{Formalism} \label{sec:formalism} 

In this section we describe the theoretical formalism for the models and estimators used. We will closely follow the notation of \cite{Singh2015,Singh2021}.
	
To model the intrinsic alignments of galaxies, we will use the linear alignment model \citep{Catelan2001,Hirata2004}. The linear alignment model states that the intrinsic shapes of galaxies are correlated with the initial tidal field at the time of the formation of galaxies. 
As a result, the correlated shear in the intrinsic galaxy shapes can be written as:
\begin{equation}
\gamma^I=(\gamma^I_+, \gamma^I_\times)=-\frac{C_1}{4\pi G}(\partial^2_x-\partial^2_y,\partial_x\partial_y)\phi_p,
\label{eqn:gamma_phi}
\end{equation}
where $\gamma^I$ is the intrinsic shear with two components $\gamma^I_+, \gamma^I_\times$, $\phi_p$ is the gravitational potential at the time of galaxy formation, \refresponse{and} $C_1$ is an unknown constant encapsulating the strength of
the linear response of galaxy shapes to the tidal fields. Later in this section, when defining the power spectra of galaxy shapes, we will use the non-linear matter power spectra instead of the linear matter power spectra as expected from the linear alignment model (a modification that was proposed by \citealt{Hirata2004} and implemented by \citealt{Bridle2007}).  The resultant model is called the non-linear alignment model (NLA).

\subsection{Power spectra and Correlation functions}
		
The correlations among different cosmological observables are typically measured and quantified using the power spectrum or its Fourier counterpart, the two-point correlation function. To study the intrinsic alignments of galaxy shapes, we measure the auto- and cross-correlations between the galaxy positions and their shapes, quantifying the statistical tendency for galaxy shapes to point towards nearby overdensities. For these measurements, the galaxy samples used for tracing the galaxy density field and the intrinsic shear field (IA) can in general be different. In this paper, we will denote the quantities associated with the density tracer sample with subscript $D$ and for the sample with shape measurements as $S$. 

In the linear alignment model, the \refresponse{primordial potential $\phi_p$ can be related to the linear overdensity field $\delta$.  We additionally adopt the linear galaxy bias model $\delta_g = b\delta$.  With these assumptions, correlations between auto- and cross-correlations of intrinsic shear and galaxy shapes can all be written in terms of the matter power spectrum $P_\delta$} as \citep[\refresponse{see section III~A of}][]{Hirata2004}		
\begin{align}
P_{gg}(\bm{k},z)&=b_D b_S P_\delta (\bm{k},z)\label{eqn:LA_gg}\\
P_{g+}(\bm{k},z)&=A_I b_D \frac{C_1\rho_{\text{crit}}\Omega_m}{D(z)} \frac{k_x^2-
k_y^2}{k^2} P_\delta (\bm{k},z)\label{eqn:LA+}\\
P_{++}(\bm{k},z)&=\left(A_I \frac{C_1\rho_{\text{crit}}\Omega_m}{D(z)} 
\frac{k_x^2-k_y^2}{k^2} \right)^2P_\delta (\bm{k},z)\label{eqn:LA++}\\
P_{g\times}(\bm{k},z)&=A_I b_D\frac{C_1\rho_{\text{crit}}\Omega_m}{D(z)}\frac{k_x 
k_y}{k^2}P_\delta(\bm{k},z)\label{eqn:LAx}.
\end{align} 
Here $P_{gg}$ is the galaxy power spectrum; $P_{g+}, P_{g\times}$ are the cross-power spectra between galaxy positions and the two components of galaxy shapes; and $P_{++}$ is the galaxy shape power spectrum. The $+$ shape component is aligned with the separation vector connecting the galaxy pair, while the $\times$ shape component is aligned at a 45$^\circ$ angle with respect to it.
$b_S, b_D$ are the linear galaxy bias parameters for the shape and density samples, respectively. By convention, $C_1\rho_{crit}$ is fixed to 0.0134 \citep{Joachimi2011} and the unknown response of galaxy shapes to tidal fields is encapsulated in the 
parameter $A_I$. $P_{\delta}$ is the matter power spectrum at the observed redshift of \refresponse{the} galaxies and is divided by the growth function, $D(z)$, for the case of the galaxy shapes, to account for the fact that the linear alignment model assumes alignments are set at the time of galaxy formation. To account for the non-linear evolution of matter clustering as in the NLA model \citep{Bridle2007}, the non-linear matter power spectrum is used as $P_{\delta}$.
		
\refresponse{Direct measurements of intrinsic alignments in real data are subject to contamination from lensing \citep[see, e.g.,][for detailed explanation of the various lensing contamination terms and their dependence on redshift quality]{Joachimi2011,Samuroff2022}.  The formalism to account for this contamination in past work was for projected correlation functions.  This contamination is not an issue for our simulated galaxy sample in IllustrisTNG, as there we have used the directly measured shapes and other properties without adding lensing contributions to number counts or shapes.  For real data it is an issue, even for samples with high-quality redshifts such as the SDSS samples used in this work, for which lensing contamination to direct IA measurements is at most a $\sim$10 per cent correction.  Therefore, we defer a direct calculation of the impact of lensing on the multipole-based estimator described below to future work, and neglect the issue of lensing contamination in this work.}
  
The correlation functions are the Fourier counterparts of the power spectra and can be obtained as 
\begin{align}
\xi_{ab}(\bm{r},z)=&\int \frac{\mathrm{d}^3k}{(2\pi)^3}P_{ab}
(\bm{k},z)e^{i(\bm{k}\cdot\bm{r})}.\label{eqn:xi}
\end{align}
Correlation functions or power spectra measured as a function of three dimensional vectors, $\bm{r}$ or $\bm{k}$, are noisy in individual bins. The bins are also redundant as the correlations only depend on amplitudes $r$ or $k$, since the universe is isotropic and homogeneous. As a result it is convenient to compress the measurements further into either the first few multipole moments, or to project them along one axis.
Next we discuss these two forms of compression.
	
\subsection{Multipoles}
	
\subsubsection{Real Space}
Since the universe is homogeneous and isotropic, the correlation function and the power spectra as measured in configuration space (no redshift space 
distortions) are
only a function of the magnitude of the wave vector (k=$|\bm{k}|$) or the separation vector ($r=|\bm{r}|$). In such a case, it is convenient to express the correlation functions in terms of their multipole moments as 
\begin{align}\label{eqn:multipole}
&\xi^{\ell,s_{ab}}_{ab}(r)=\frac{2\ell+1}{2} \frac{(\ell-s_{ab})!}{(\ell+s_{ab})!} \int \mathrm{d}\mu_r\, L^{\ell,{s_{ab}}}(\mu_r)\xi_{ab}(r,\mu_r)\\
&\mu_r = \frac{\Pi}{r}; \mu_k =\frac{k_{z}}{k}\label{eqn:mu}
\end{align}
where $(\mu_r, \mu_k)$ are the cosine of the angle between $(\bm{r}, \bm{k})$ and the line of sight direction. $\Pi$ is the line of sight component of 
$\bm{r}$ and the component in the plane of the sky will be denoted as $r_\mathrm{p}$. 
$ L^{\ell,{s_{ab}}}$ are associated Legendre polynomials of order $\ell$ and $s_{ab}$, and are non-zero only when $\ell\geq s_{ab}$.  
Analogous expansions can be done for power spectra measurements as well, including the $E$/$B$ mode power spectra for intrinsic alignments\refresponse{. Measurements of $E$/$B$ mode power spectra for IA in sims have been presented by \citet{Kurita2021}, and \citet{2023PhRvD.108h3533K} also carried out a multipole expansion of an IA power spectrum in real data.}

For the galaxy correlation function in real space, $s_{ab}=0$ (the galaxy density field is spin zero), and because of symmetry, only the monopole ($\ell=0$) term exists (we are assuming no redshift space distortions in this section).  
For the case  of galaxy shapes, as was discussed in \refresponse{section~2.3 of} \cite{Singh2016ia}, we measure the shapes projected onto the plane of the sky, which introduces an additional anisotropy term of the form $1-\mu^2$ (both in real space $\mu_r$ and Fourier space, $\mu_k$). 
\refresponse{Functionally this term is introduced for each factor of galaxy shapes, meaning it appears once in the $g+$ correlations and is squared in the $++$ correlations.  As a result, when defining the multipoles using Legendre polynomials which act as a `matched filter', the matched filters for \xigp\ and \xipp\ will be  the associated Legendre polynomials, 
$ L^{2,2}\propto(1-\mu^2)$ and $ L^{4,4}\propto(1-\mu^2)^2$,  respectively.} Mathematically, the galaxy shapes or the tidal fields are 
spin two objects, since they involve two factors of spin raising operators and are symmetric under rotation by 180$^\circ$. This implies that $s_{ab}=2$ for \xigp\ and $s_{ab}=4$ for \xipp\ (given that there are two shapes in the auto-correlation).
	      
The NLA model predictions for correlation function multipoles can be obtained through Hankel transforms of power spectra as
\begin{equation}
\xi_{ab}^{\ell,s_{ab}}(r)=(-1)^{\ell/2}\alpha_{\ell,ab}\frac{C_{ab}}{2\pi^2}\int \mathrm{d}k \, k^2 P_{\delta}(k)j_{\ell}(kr)
\label{eq:xi_multipole_model}
\end{equation}
where $j_{\ell}$ are the spherical bessel functions, $\alpha_{\ell,ab}$ are constants (defined in the next section) that depend on the observables and the amplitude of the associated Legendre polynomials; and $C_{ab}$ are the
observable- and cosmology-dependent constants defined in Eqs.~\eqref{eqn:LA_gg}--\eqref{eqn:LAx}, which relate the matter power spectrum to the power spectra of the observables. 
		
We transform the matter power spectrum predictions from the model to the correlation function multipoles via the Hankel transforms implemented in 
the \texttt{mcfit} package \citep{Li2019}.

\subsubsection{Redshift space}
Since line-of-sight distances to galaxies are estimated using redshifts, galaxy velocities introduce coherent distortions in the estimated distances,  
which induce an anisotropy in the correlation functions or power spectra. At the linear order, this anisotropy can be modeled using a Kaiser 
correction factor \citep{Kaiser1987}. Under the Kaiser model, the galaxy over-density field ($\delta_g$) in redshift space is given as
\begin{equation}
\delta_g(k, \mu_k) = b_g\delta_m + f\mu_k^2 \delta_m = b_g\delta_m (1+\beta_g\mu_k^2)
\end{equation}
where $b_g$ is the galaxy bias, $f$ is the linear growth rate of the matter density perturbation, and $\beta_g=f/b_g$.
In this case, the galaxy-galaxy and galaxy-shape cross-correlation functions are given as 
\begin{align}
P_{gg}(\bm{k},z)&=b_D b_S P_\delta (\bm{k},z)(1+\beta_S\mu_k^2)(1+\beta_D\mu_k^2)\label{eqn:LA_gg_rsd}\\
P_{g+}(\bm{k},z)&=A_I b_D \frac{C_1\rho_{\text{crit}}\Omega_m}{D(z)} \frac{k_x^2-
k_y^2}{k^2} P_\delta (\bm{k},z)(1+\beta_D\mu_k^2).
\end{align} 
		    
As in the previous subsection, we can write the correlation functions and power spectra in redshift space in terms of their multipole moments. The additional factors of $(1+\beta_g\mu_k^2)$, where $g$ refers to galaxy positions from either the $S$ or $D$ samples,  
result in additional quadrupole ($L^{2,0}$) and 
hexadecapole $L^{4,0}$ terms for galaxy auto-correlations, while the 
galaxy-shape cross correlation functions gain an additional hexadecapole ($L^{4,2}$) contribution.

As described in Eq.~\eqref{eq:xi_multipole_model}, the multipoles of the correlation functions can be written in terms of the Hankel transforms of the power spectrum. 
The coefficients $\alpha_{\ell,gg}$ for the monopole ($\ell=0$), quadrupole ($\ell=2$) and hexadecapole ($\ell=4$) of the galaxy auto-correlations are given by 
\citep{Baldauf2010,Singh2021}
\begin{align}
\alpha_{0,gg}&=1+\frac{1}{3}(\beta_S+\beta_D)+\frac{1}{5}\beta_S\beta_D\label{eq:rsd_alph0}\\
\alpha_{2,gg}&=\frac{2}{3}(\beta_S+\beta_D)+\frac{4}{7}\beta_S\beta_D\label{eq:rsd_alph2}\\
\alpha_{4,gg}&=\frac{8}{35}\beta_S\beta_D.
\end{align}	      
	
The coefficients $\alpha_{\ell,g+}$ for the quadrupole ($\ell=2$) and hexadecapole ($\ell=4$) of the galaxy-shape cross-correlations are given 
by 
\begin{align}
\alpha_{2,g+}&=\frac{1}{3}\left(1+\frac{1}{7}\beta_D\right)
    \\
\alpha_{4,g+}&=\frac{2}{105}\beta_D\label{eq:rsd_alph4}
\end{align}

\subsubsection{Wedges}

On small scales, our model  does not include the effects of non-linear clustering \citep[beyond the Halofit correction;][]{Smith2003,Takahashi2012} or  
non-linear effects in the galaxy velocities, such as the Fingers-of-God effect. Because of these issues, modeling the multipole moments down to very small scales
can results in biased estimates of model parameters, and it is desirable to remove the small scale information from our analysis. 
		 
The Fingers-of-God effect caused by the motion of satellite galaxies within dark matter halos smears the correlation function along the line of sight, but is localized to scales
that are comparable to the size of the largest halos in our sample. Thus it is desirable to remove the scales that lie within a few times the typical halo 
size along the plane of the sky.  We remove these scales by applying cuts on $r_\mathrm{p}$, which is the projection of the separation vector $\bm{r}$ onto the plane of the sky. This 
results in a wedged measurement of the correlation function. The wedged multipoles are measured by modifying Eq.~\eqref{eqn:multipole} to 
\begin{align}\label{eqn:multipole_wedge}
&\widetilde{\xi}^{\ell,s_{ab}}_{ab}(r)=\frac{2\ell+1}{2} \frac{(\ell-s_{ab})!}{(\ell+s_{ab})!} \int \mathrm{d}\mu_r\, \Theta(r_\mathrm{p}) L^{\ell,{s_{ab}}}(\mu_r)\xi_{ab}(r,\mu_r).
\end{align}
We introduced the scale cut operator $\Theta(r_\mathrm{p})$, which is zero below a chosen $r_\mathrm{p}$ value, $r_\mathrm{p}<{r_{\mathrm{p},\text{min}}}$, and one otherwise.
For the theory predictions, we transform the model for $\xi^{\ell,s_{ab}}(r)$  in Eq.~\eqref{eq:xi_multipole_model} to $\xi^{\ell,s_{ab}}(r,\mu_r)$ by multiplying it with
$L^{\ell,s_ab}(\mu_r)$, and then transforming it back using Eq.~\eqref{eqn:multipole_wedge}. The $r_p>r_{p,\text{min}}$ cut also implies that 
$r>r_{p,\text{min}}$, thereby removing the majority of the scales that are most affected by non-linear clustering effects.

Note that our definition of wedges is slightly different from the more common definitions in the literature \citep[e.g.,][]{Kazin2013,Sanchez2017}. Typically a 
fixed cut is applied on $\mu$,  
such that only the $\mu$ bins defined within a certain range are used to compute the wedges. We are, however, applying the 
cuts on $r_\mathrm{p}$, which results in a variable range of $\mu$ for different $r$ bins. Since our goal is to remove most of the effects of non-linear galaxy velocities,
cuts in $r_\mathrm{p}$ allow us to remove the most contaminated bins while retaining most of the other bins, resulting in a more optimal estimator for our purposes.
	
\subsection{Projected correlation functions}
We also measure the projected correlation functions to compare against our multipole measurements. The model for the projected correlation functions can 
be obtained from the multipoles as \citep{Baldauf2010,Singh2021}
\begin{equation}
w_{ab}(r_\mathrm{p})=\sum_{\ell} 2
\int_0^{\Pi_\text{max}}\mathrm{d}\Pi\,\xi_{ab}^{\ell,s_{ab}}(r) L^{\ell,s_{ab}}\left(\frac{\Pi}{r}\right).
\label{eq:w_model}
\end{equation}
The computation in Eq.~\eqref{eq:w_model} is more accurate compared to the commonly used projected Hankel transform of the matter power spectrum using Bessel 
functions, as Eq.~\eqref{eq:w_model}  models the residual effects of redshift space distortions more accurately.
			
The projected density-shape correlation function, \wgp, is typically used to measure and quantify the IA signal. The \wgp\ estimator has the advantage that it is relatively easy to relate 
to the weak gravitational lensing measurements, which are sensitive to the projected modes. Projected correlation functions are also relatively less sensitive to the 
redshift space distortion effects from galaxy velocities and/or less accurate and noisy photometric redshifts. However, \refresponse{achieving these gains} requires the use of a large $\Pi_\text{max}$ value \refresponse{(see figure~8 in \citealt{Baldauf2010})}, which 
also increases the noise. The multipole moments, on the other hand, use the information more optimally, at least on the scales where RSD can be modeled well. 
 As we will show later in this work, this leads to tighter constraints on IA model parameters, which is more useful for constraining parametric models used for marginalizing over intrinsic alignments in weak gravitational lensing measurements (the exception to this will be IA self-calibration methods, such as in \citealt{Yao2019}). 

\section{Data}\label{sec:data}

\refresponse{Here we describe the simulated and real datasets used for this work.  Use of simulated data allows for controlled tests separating out different physical effects in a way that cannot be done in real data.  Use of real data, in particular a dataset for which intrinsic alignments have already been well studied, is important to confirm the quantitative improvements in performance of the new estimator in reality.}

	\subsection{Simulated data: IllustrisTNG}

  The IllustrisTNG suite of cosmological simulations includes magneto-hydrodynamical galaxy physics in periodic boxes of $\sim$ 50, 100, 300 Mpc  each with three different varying resolutions \citep{{ tng-bimodal,pillepich2018illustristng, Springel2017illustristng, Naiman2018illustristng, Marinacci2017illustristng,tng-publicdata}}. All of the simulations were run using the moving mesh code \textsc{Arepo} \citep{arepo} using cosmological parameters from Planck CMB measurements under the assumption of a flat $\Lambda$CDM model \citep{planck2016}. 
  
  In this study we use the IllustrisTNG100 dataset due to its balance between large volume and high resolution. 
  This particular simulation has $2 \times 1820^3$ resolution elements with a gravitational softening length of 0.7~\refresponse{$h^{-1}$}kpc  for dark matter and star particles with  masses of $7.46 \times 10^6 M_\odot$ and $1.39 \times 10^6 M_\odot$, respectively. In addition to the galaxy formation and evolution physics, the simulations also include  radiative gas cooling and heating; star formation; stellar evolution with supernovae feedback;
 AGN and blackhole feedback \citep[for more details, see][]{tng-methods,tng-agn}.
 The friends-of-friends algorithm \citep[FoF;][]{fof} and the \textsc{subfind}  \citep{subfind} algorithms were employed to identify halos and subhalos in the simulations, respectively. 

The latest snapshot at redshift zero was used for our analysis.  A mass cut of $ \log_{10}(M_*/M_\odot) \ge 10$ was enforced in order to obtain reliable galaxy shape measurements and to avoid resolution limitations \citep{jagvaral-gal-decomp}. The galaxy shapes were measured using the simple inertia tensor using all particles in the subhalo  \citep[for more information, please refer to][]{jagvaral-bulge-disc-ia}.  
Further, since bulge-dominated (elliptical) galaxies exhibit higher IA signal (and hence higher signal-to-noise) as shown in that paper, we selected the  elliptical sample using method described in \cite{jagvaral-gal-decomp}. To summarize the galaxy morphological decomposition method: using the dynamics of the star particles, we calculate the disc-to-total mass ratio for each galaxy, then assign the lower 33rd percentile to be the elliptical sample.  The selected sample contains approximately 2000 galaxies. 

\refresponse{Our previous work cited in this section shows some properties of this simulated galaxy sample.  For example,  figure~3 of \citet{jagvaral-bulge-disc-ia} shows that it has a shape distribution comparable to that of elliptical galaxies in real data, suggesting that its morphology is realistic. Its luminosity is lower on average than that of typical LRGs (such as in the BOSS LOWZ sample).  However, for the purpose of this work, it does not matter whether the sample is comparable to the samples we measure in real data -- in fact, showing that the multipole-based estimator is more efficient than projected statistics on two quite different samples bodes well for its utility on a variety of samples in real data.}

	\subsection{Real data: SDSS}

The Sloan Digital Sky Survey  \citep[SDSS;][]{2000AJ....120.1579Y} encompassed both multi-band drift-scanning imaging \citep{1996AJ....111.1748F, 1998AJ....116.3040G, 2001AJ....122.2129H, 2002AJ....123.2121S,2004AN....325..583I} and follow-up spectroscopy \citep{2001AJ....122.2267E,2002AJ....123.2945R,2002AJ....124.1810S} covering roughly $\pi$ steradians
of the sky, using the SDSS Telescope \citep{Gunn2006}.  The data were processed by automated calibration and measurement pipelines \citep{2001ASPC..238..269L,2003AJ....125.1559P,2006AN....327..821T}. This work uses data through the seventh SDSS-I/II data release \citep{2009ApJS..182..543A}, though with improved data reduction pipelines from the eighth release as part of SDSS-III  \citep{2008ApJ...674.1217P,2011ApJS..193...29A}.
  
The BOSS survey selected galaxies from the imaging for spectroscopic observations \citep{Ahn:2012,Dawson:2013,Smee:2013}. For details of the targeting algorithms and spectroscopic processing, see \citet{Bolton:2012,2016MNRAS.455.1553R}. 
For this work, we use SDSS-III BOSS Data Release 12 (DR12) LOWZ galaxies in the redshift range $0.16<z<0.36$. The LOWZ sample consists of Luminous Red Galaxies (LRGs) at $z<0.4$. The sample is approximately volume limited in the redshift range $0.16<z<0.36$, with a number density of $\bar{n}\sim 3\times10^{-4}h^3\text{Mpc}^{-3}$ \citep{Manera2015}. 
  
We combine the spectroscopic redshifts from BOSS with galaxy shape measurements from \cite{Reyes2012} to measure intrinsic alignments. \refresponse{\cite{Reyes2012} and multiple subsequent papers \citep[e.g.,][]{Mandelbaum2013,Singh2016ia} carried out an extensive suite of validation and null tests to ensure the shape measurements have sufficient quality for measurements of two-point correlation functions involving galaxy shapes.} When including \refresponse{redshift cuts $0.16< z< 0.36$ and extinction cuts $A_r\le 0.2$}, our sample includes 225181 galaxies for our LOWZ density 
sample, of which there are good shape measurements for 181933 galaxies. 

Additional \refresponse{galaxy properties have also} been computed and used when studying the LOWZ sample intrinsic alignments, such as rest-frame colours and luminosities; we refer the interested reader to \citet{Singh2015,Singh2016ia} for details.

\section{Methods}\label{sec:methods}


\subsection{Measuring correlation functions}


  The standard Landy-Szalay estimator was used for the clustering two-point correlation function  \citep{landy}:
 
\begin{equation}\label{eqn:xigg_est}
\xi_{gg}(r_{\rm p}, \Pi) = \frac{SD- DR_S - SR_D + R_SR_D}{R_SR_D},
\end{equation}
where $SD$, $R_SR_D$ and $DR_S$ are weighted counts of galaxy-galaxy, random-random and galaxy-random pairs (with $S$ and $D$ denoting shape and density samples), binned based on their perpendicular and line-of-sight separation, $r_\mathrm{p}$ and $\Pi$. 

In the periodic simulation box, the $DR$ and $RR$ values are computed analytically, and for a particular bin in $(r_\mathrm{p}, \Pi)$ are given by

\begin{align}
DR_S=R_SR_D=SR_D=\frac{N_{g,S}N_{g,D}\times\text{bin volume}}{\text{simulation volume}},
\end{align}
where $N_{g}$ is the number of galaxies \refresponse{of the desired type (density or shape sample) in the simulation}.
Since we use jackknife methods to obtain the covariance, the $DR$ and $RR$ values are rescaled by $N_\text{jk}-1$, where $N_\text{jk}$ is the number of jackknife regions. To ensure that the expected number of $DD$ pairs in the absence of clustering matches the analytical calculation, we also modify the jackknife implementation. For a galaxy pair $(i,j)$, if 
both galaxies belong to jackknife region $k$, the pair is excluded when region $k$ is excluded. However, for a cross pair where only one of the galaxies 
belongs to region $k$, the pair is only excluded with probability $50\%$. We have confirmed that the measurements obtained using this method are consistent 
with the implementation where random points were used to compute the $DR$ and $RR$ values and a simple jackknife estimator was used where pair $(i,j)$ is always 
excluded when either of the galaxies is not in region $k$.

A modified version of the Landy-Szalay estimator \citep{mandelbaum-2011,Singh2017cov}  was used for measuring  the cross-correlation function of galaxy positions and intrinsic ellipticities 
as a function of $r_\mathrm{p}$ and $\Pi$:
\begin{equation}\label{eqn:xigp_est}
\xi_{g+} (r_\mathrm{p}, \Pi) = \frac{S_+D - S_+R}{RR}.
\end{equation}
The  shape-shape correlation was estimated similarly
\begin{equation}
\xi_{++} (r_\mathrm{p}, \Pi) = \frac{S_+S_+}{RR}.
\end{equation}
where 
\begin{equation}
S_+D \equiv \frac{1}{2} \sum_{\alpha\neq \beta} w_\alpha w_\beta \, e_{+}(\beta|\alpha),
\end{equation}
\begin{equation}
S_+S_+ \equiv \frac{1}{4} \sum_{\alpha\neq \beta}    w_\alpha w_\beta \,  e_{+}(\alpha|\beta) e_{+}(\beta|\alpha),
\end{equation}
 represent the shape correlations,  
   $e_{+}(\beta|\alpha)$ is the $+$ component of the ellipticity of galaxy $\beta$ (from the shape sample) measured relative to the direction of galaxy $\alpha$ (from the density tracer sample) and $w_\alpha$ ($w_\beta)$ are weights associated with galaxy $\alpha$ ($\beta$).

 The projected two-point correlation functions (Eq.~\ref{eq:w_model}) are approximated as sums over   the line-of-sight separation ($\Pi$) bins:
 \begin{equation}\label{eq:xi_to_w_sum}
w_{ab} (r_\mathrm{p}) = \sum_{-\Pi_\mathrm{max}}^{\Pi_\mathrm{max}} \Delta\Pi \,\xi_{ab} (r_\mathrm{p}, \Pi),
\end{equation}
where  $a,b\in(g,+)$. 

For the SDSS data, we use a value of $\Pi_\text{max}=100\mpch$ \refresponse{as suggested by \citet{Baldauf2010}}. 
In the simulation, due to the limited size of the simulation box, two cuts were applied. First, a $\Pi_\text{max}$ value of 20~\mpch \, was used for both the simulation measurement and the model prediction.  Second, in order to exclude modes that are not present in the periodic box of the simulation,
a  $k_\text{min}$ 
cutoff of $\pi/L_\text{box} = 0.04$ h/Mpc was used for the model predictions.  

To compute the multipole moments, we measure the correlation functions in the $(r,\mu_r)$ coordinate system to obtain $\xi(r,\mu_r)$ using the estimators in 
Eqs.~\eqref{eqn:xigg_est} and~\eqref{eqn:xigp_est}. Then  we use Eq.~\eqref{eqn:multipole_wedge} to 
compute the wedged multipole moments.

\subsection{Covariance matrices}

The covariance matrices for our measurements are estimated using the jackknife method.  First, we divide the simulation box or the SDSS survey  into $N_\mathrm{jk}$ equal-volume regions, where $N_\text{jk}=100$ for SDSS and $N_\text{jk}=49$ for the simulation box.  Then we perform the measurements $N_\mathrm{jk}$ times, while dropping one jackknife region during each measurement, such that we are performing the measurement for $N_\mathrm{jk} -1 $ regions. Given a data vector $\psi_i \in (\xi, w)$ -- with $i$ indexing the bins in our measurement, for example in $(r,\mu_r)$ space or in $(r_\mathrm{p},\Pi)$ space -- the covariance matrix is computed as:
\begin{equation}
   C^\mathrm{jk}_{ij}  = \frac{N_\mathrm{jk}-1}{N_\mathrm{jk}} \sum^{N_\mathrm{jk}}_{m=1} \big( \psi^m_i - \overline{\psi}_i \big) \big( \psi^m_j - \overline{\psi}_j \big)
\end{equation}
where $\overline{\psi}_i = \frac{1}{N_\mathrm{jk}} \sum^{N_\mathrm{jk}}_{m=1} \psi^m_i $.  Here $m$ is indexing the jackknife regions.
	\begin{figure*}
		\centering
	\begin{subfigure}[t]{0.49\textwidth}
		\centering
		\includegraphics[width=\columnwidth]{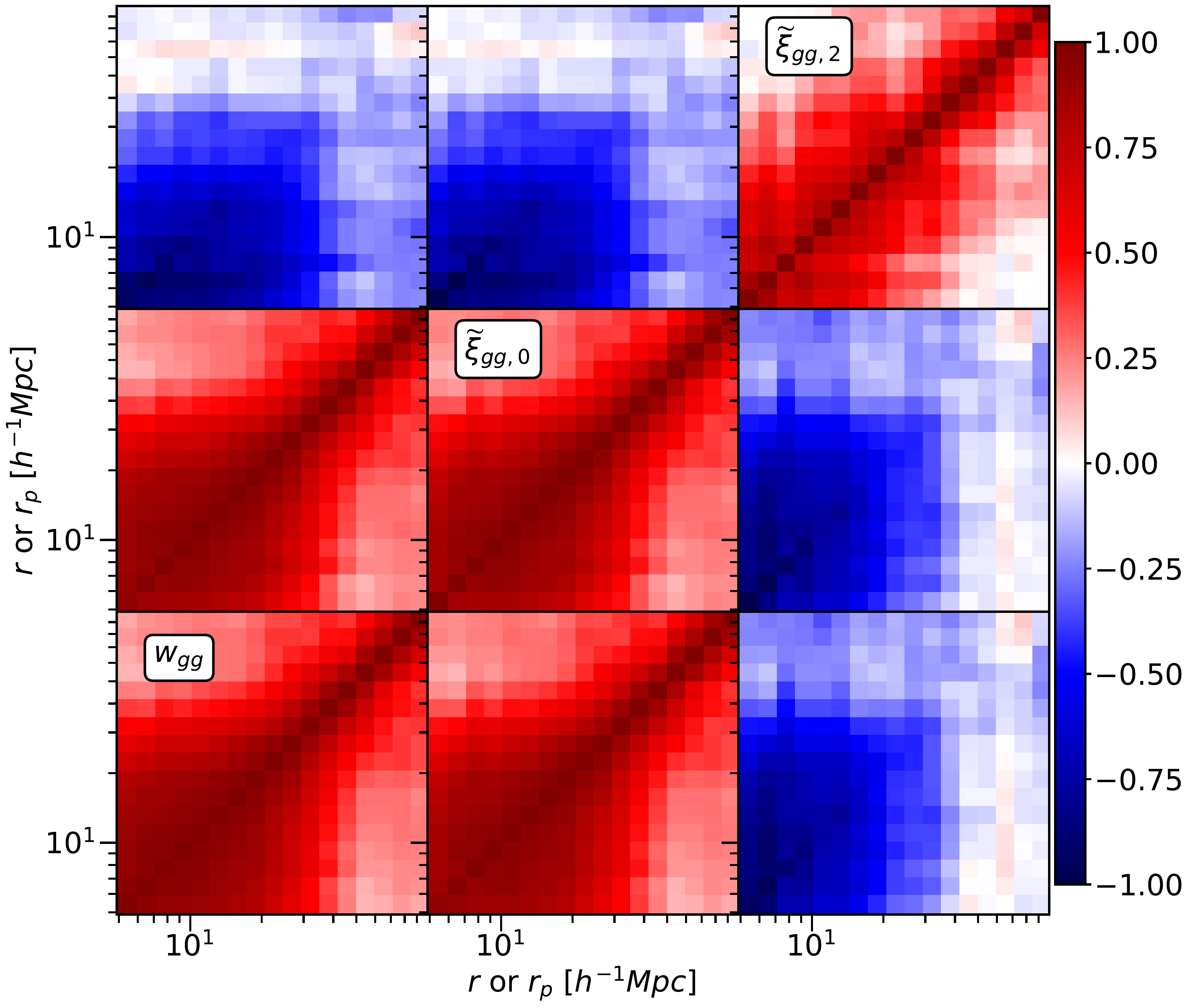}
		\caption{}
		\label{fig:corr_gg}
	\end{subfigure}
	\begin{subfigure}[t]{0.49\textwidth}
		\centering
		\includegraphics[width=\columnwidth]{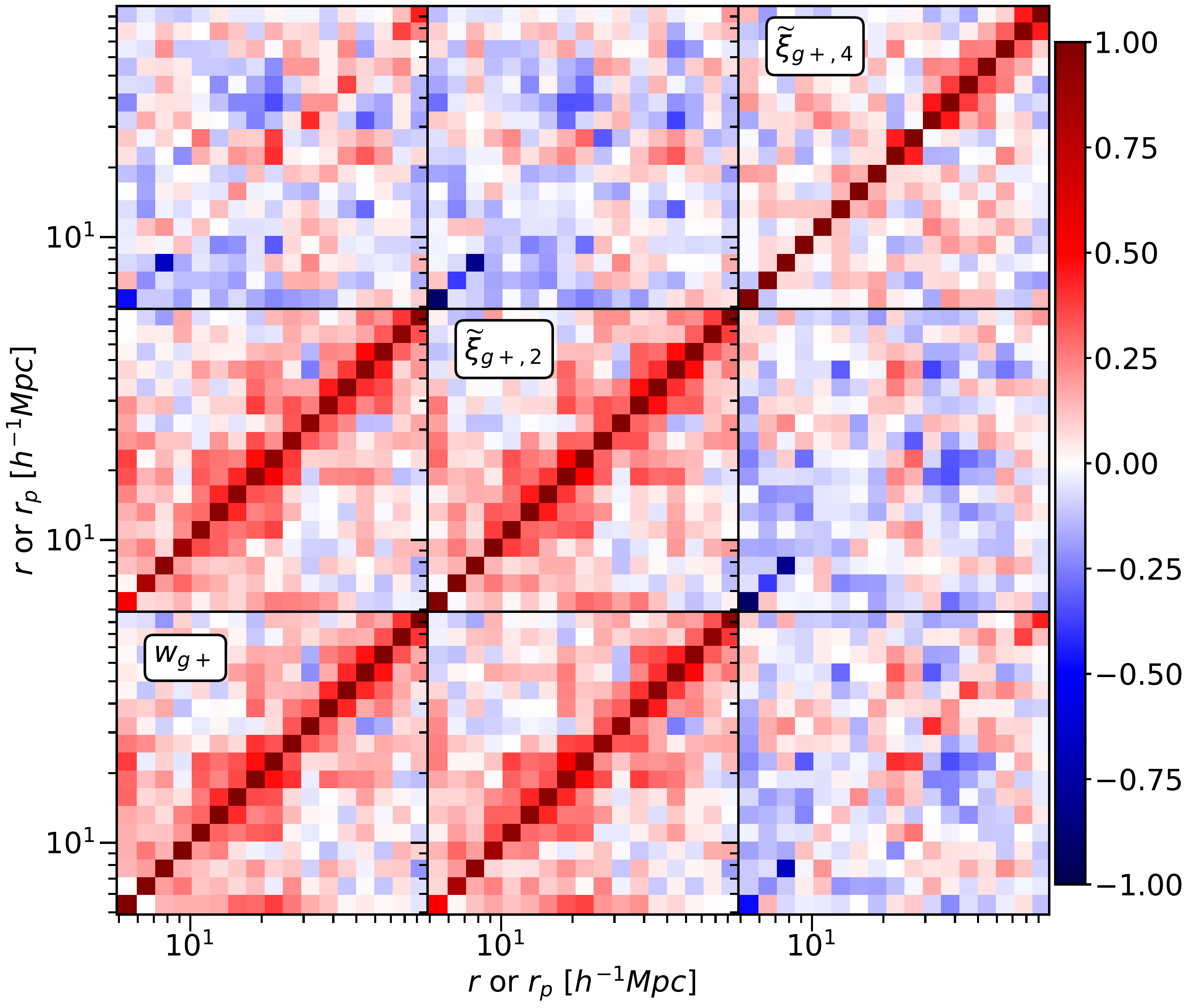}
		\caption{}
		\label{fig:corr_gg}
	\end{subfigure}
	\caption{\refresponse{Correlation matrices for a) galaxy clustering and b) IA estimators applied to the BOSS LOWZ sample. We used $r_p>5\mpch$ for projected correlation functions and multipoles wedges.} \label{fig:correlationmatrices}
	}
	\end{figure*}

\refresponse{As an illustration, Fig.~\ref{fig:correlationmatrices} shows the correlation matrices resulting from this exercise for the BOSS LOWZ sample.  As shown, the $gg$ correlation functions have significant off-diagonal elements in their covariance matrices because they are cosmic variance dominated, while the $g+$ correlations are much closer to diagonal because they are shape noise dominated.}

While jackknife covariances can be noisy, they have been shown to be consistent with covariances estimated based on numerical calculations  \citep{Singh2017cov}. Typically, the noise effects in the inverse covariance
are approximately corrected via a Hartlap correction \citep{Hartlap2007}. However, as described in the next section, we will be fitting the models to each of the jackknife 
measurements and then directly obtaining the jackknife mean and variance for the parameters. In this case, the Hartlap correction is unnecessary since 
it is a simple rescaling of the covariance and hence of the loss function, which does not impact our fitting procedure. The noise in the covariance still results in
suboptimal constraints -- i.e., the noise in our parameter constraints is likely to be over-estimated. Such limitations unfortunately exist for all covariance estimators and corrections for these effects is outside the scope of this work. 

\subsection{Fitting to models}

We only fit for two parameters in our models\refresponse{:} the galaxy bias $b_g$ and the IA amplitude $A_I$. To obtain the parameters and their uncertainties we fit the models to the measurements in each jackknife region separately and then obtain their jackknife mean and variance as defined in previous section. The fits in each region are obtained via $\chi^2$ minimization, where $\chi^2$ is defined as 
\begin{align}
    \chi^2 = &(\psi_{gg}-\psi_{gg,\text{model}})^T \mathsf{C}^{-1}_{(gg)}(\psi_{gg}-\psi_{gg,\text{model}}) \nonumber \\+ 
    &(\psi_{g+}-\psi_{g+,\text{model}})^T \mathsf{C}^{-1}_{(g+)}(\psi_{g+}-\psi_{g+,\text{model}})
\end{align}
where $\mathsf{C}_{(gg)}$ and $\mathsf{C}_{(g+)}$ are the covariances for the auto-correlations or density-shape cross-correlations obtained via the jackknife method, and $\psi$ represents either the multipole or projected correlation function as in the previous subsection. Since we have a limited number of jackknife regions, we do not use the full covariance between $gg$ and $g+$ to limit the effects of noise in the covariance (we omit the cross covariance term). This should not have a large impact on our analysis, as the covariances are noise dominated and hence the cross covariance between $gg$ and $g+$ measurements is expected to be small.
 
\section{Results}\label{sec:results}

In this section we present and interpret the measurements using the SDSS-III BOSS LOWZ sample and using the Illustris-TNG simulations. 

 	\subsection{Results from SDSS data}
	
\begin{figure*}
\begin{subfigure}[t]{\columnwidth}
\centering
\includegraphics[width=\columnwidth]{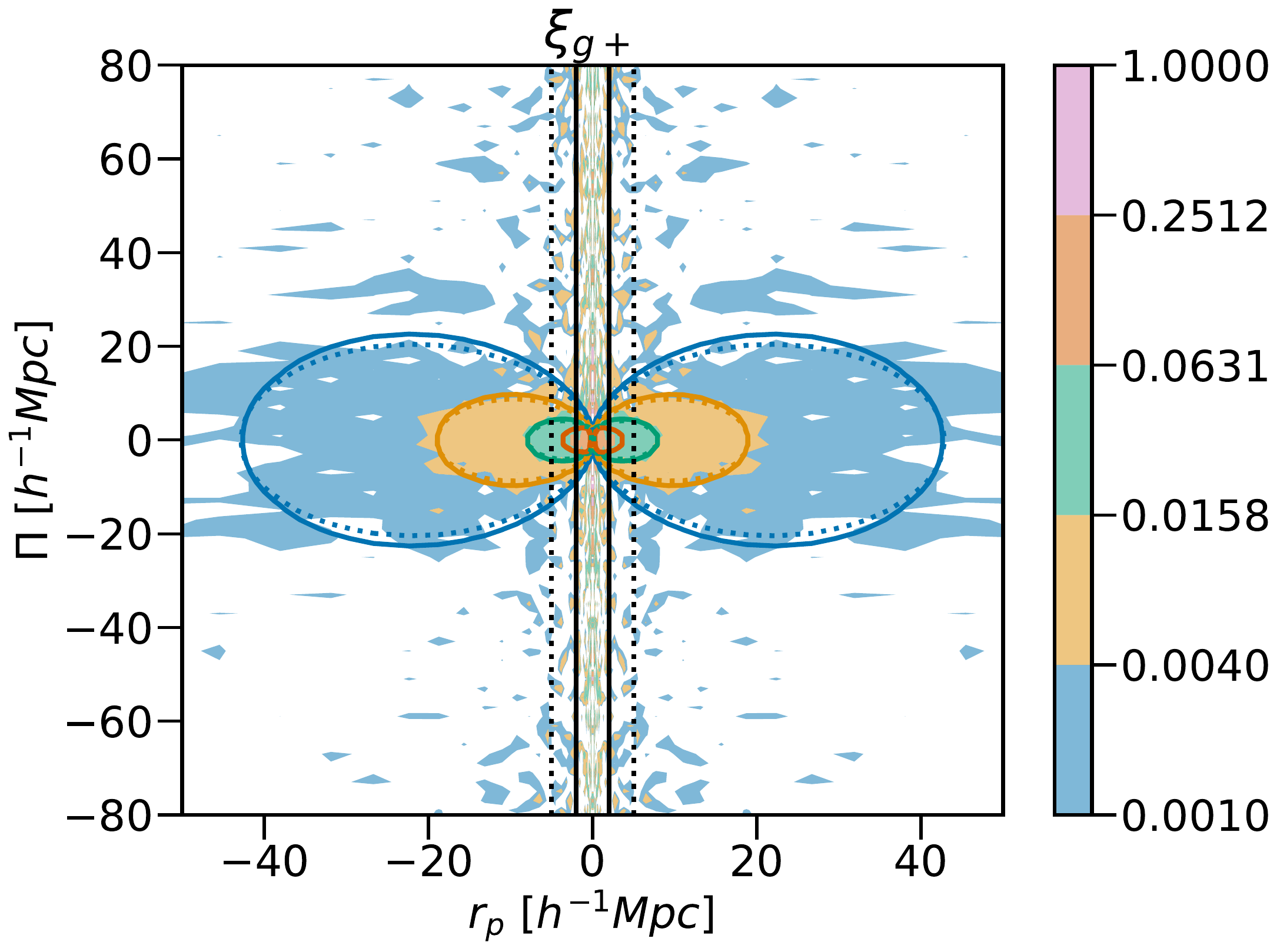}				
\caption{}
\label{fig:xigp_2d}
\end{subfigure}
\begin{subfigure}[t]{\columnwidth}
\includegraphics[width=\columnwidth]{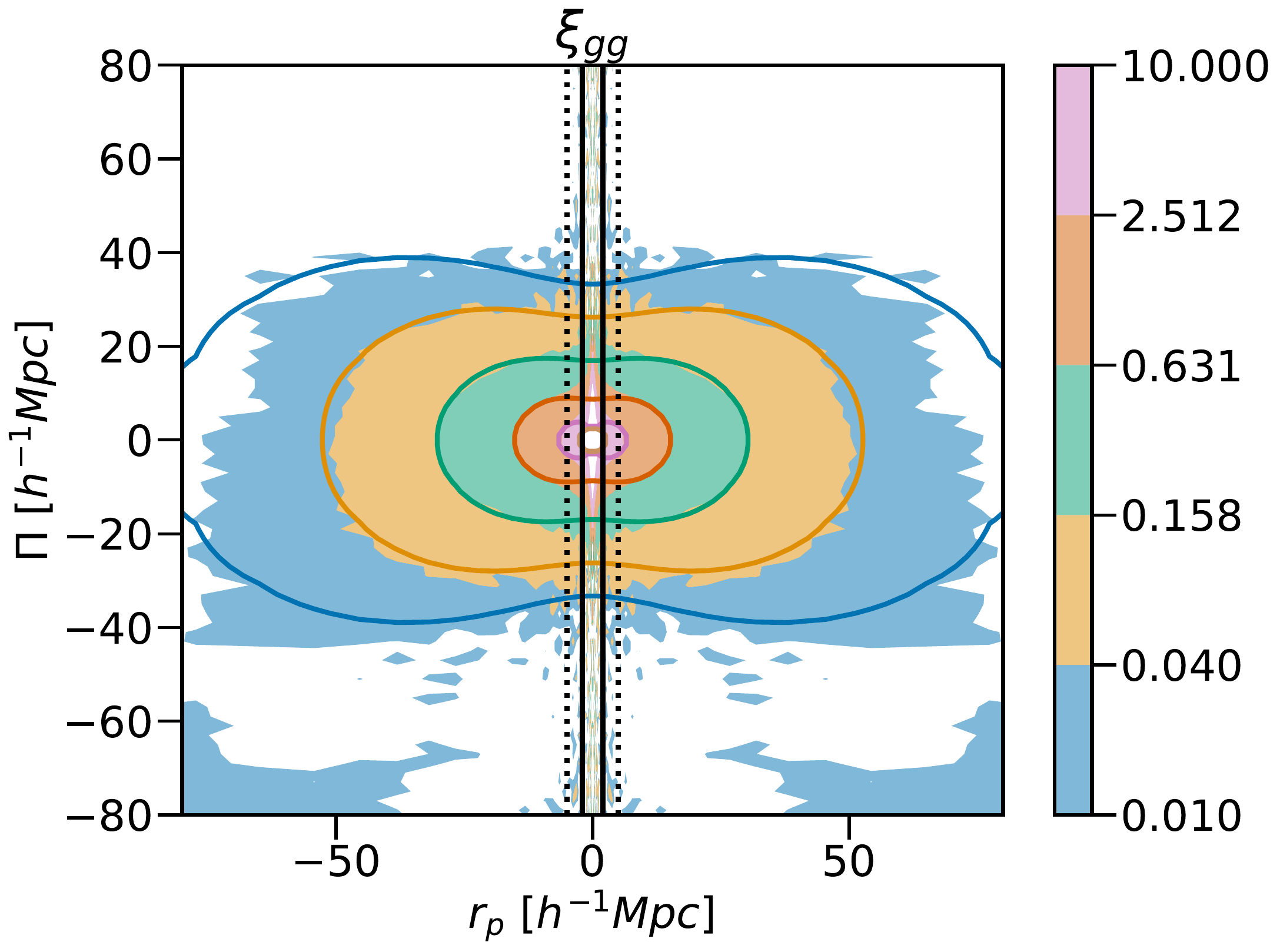}
 \caption{}
\label{fig:xigg_2d}
\end{subfigure}
\caption{ Galaxy-shape (Fig.~\ref{fig:xigp_2d}) and galaxy-galaxy (Fig.~\ref{fig:xigg_2d}) correlation function measurements for the full LOWZ 
galaxy sample in the $(r_\mathrm{p},\Pi)$ space. The solid contours show the measurements from the data, the solid lines are the model in 
redshift space (with the Kaiser correction, which is intended to be valid on large scales), and dotted lines are the model without redshift space 
correction (shown only for \xigp). 
We see that the model appears to be consistent with the data on large scales ($r_\mathrm{p}\gtrsim5\mpch)$ within the statistical 
uncertainties. On scales below a few \refresponse{$h^{-1}$}Mpc, there are considerable deviations, especially due to the 
Fingers-of-God effect (FoG), which is not captured in our models. The vertical solid (dotted) black lines are placed at $r_\mathrm{p}=2\mpch$ 
($5\mpch$). The region with
$r_\mathrm{p}\le 5\mpch$ contains almost all of the FoG effect, and hence we use $r_\mathrm{p}>5\mpch$ when defining the wedged 
multipole moments from the data and model.
 By convention $r_\mathrm{p}$ is defined to be positive, but the plot has been made with mirror images about $r_\mathrm{p}=0$.
 }
\label{fig:corr_2d}
\end{figure*}

\begin{figure*}
\begin{subfigure}[t]{\columnwidth}
\centering
\includegraphics[width=\columnwidth]{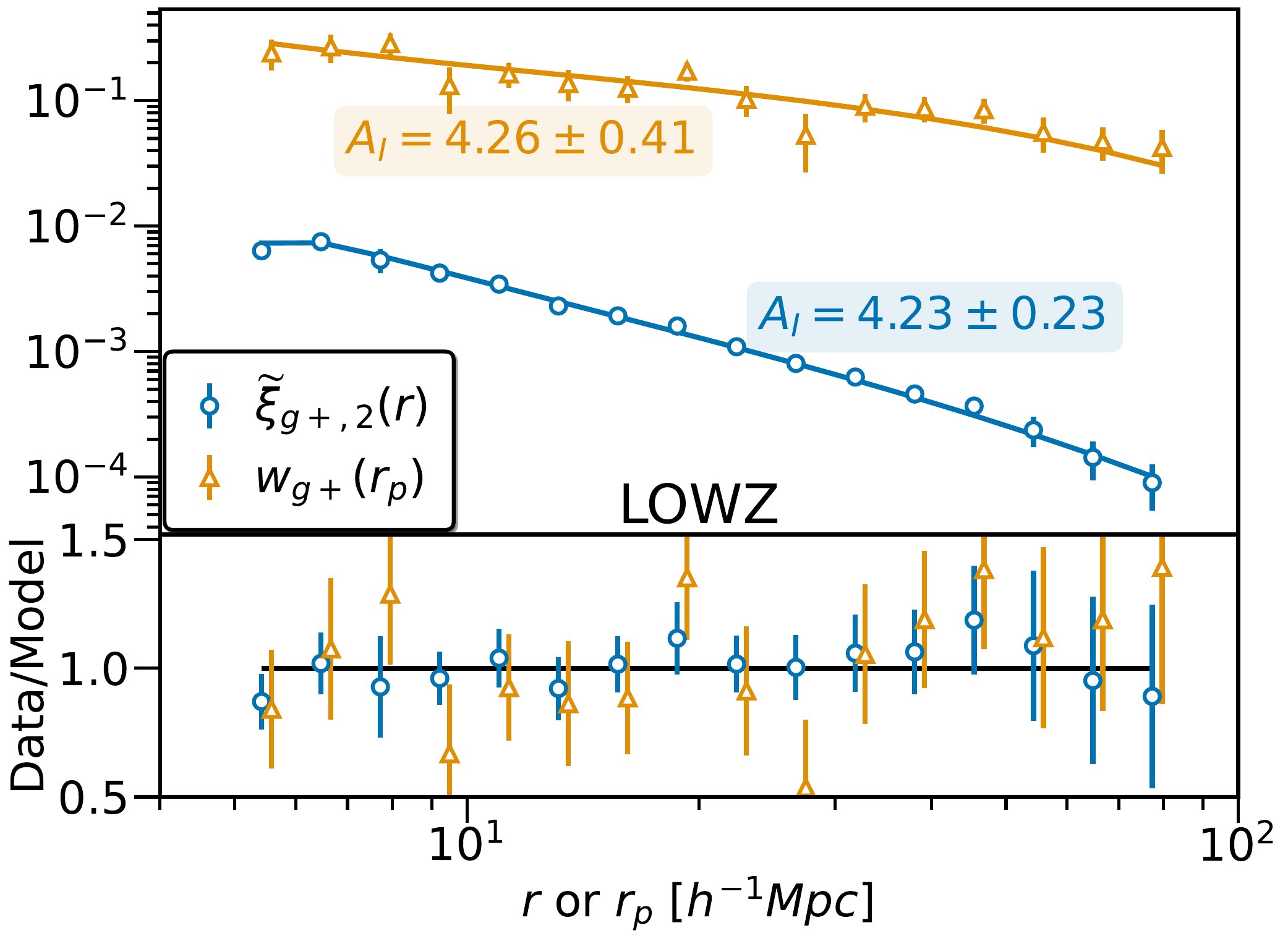}\caption{}
\label{fig:xigp_fit}
\end{subfigure}
\begin{subfigure}[t]{\columnwidth}
\includegraphics[width=\columnwidth]{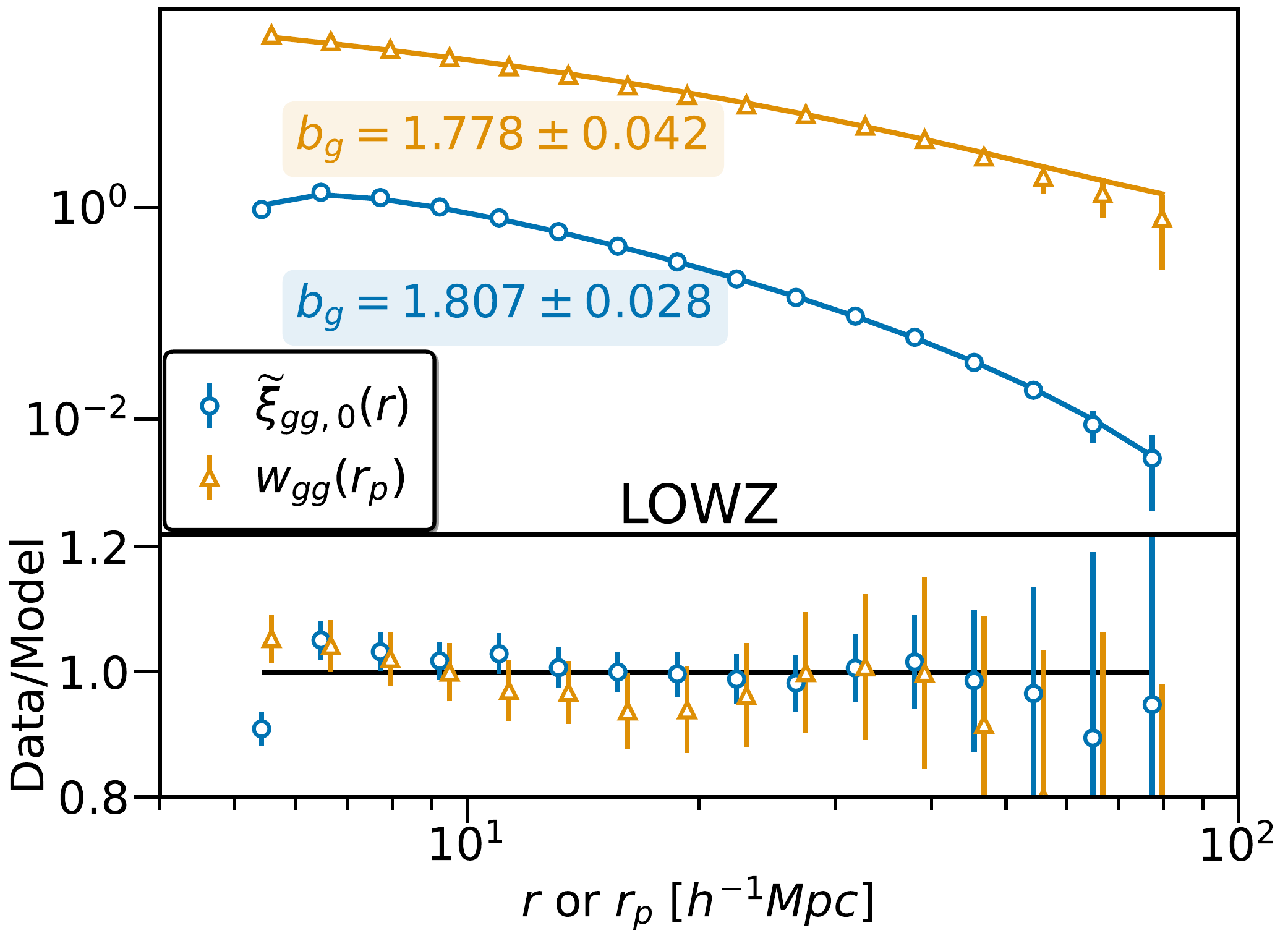}
\caption{}
\label{fig:xigg_fit}
\end{subfigure}
\caption{For the full LOWZ sample, we show \wgp\ and the $\widetilde{\xi}_{g+}$  wedged quadrupole  (Fig.~\ref{fig:xigp_fit}), and \wgg\ and the 
$\widetilde{\xi}_{gg} $ wedged monopole   (Fig.~\ref{fig:xigg_fit}) measurements. The wedged multipoles use a  $r_\mathrm{p}>5\mpch$ cut. The solid points 
show the measurements from data with jackknife errorbars while the solid lines show the model predictions. The lower panel in both figures shows the ratio 
of data measurements to the model. 
The model is in excellent agreement with the data on all scales shown, given that we have removed the scales below $r_\mathrm{p}=5\mpch$, as they are 
affected by non-linearities introduced by the Fingers-of-God effect and non-linear clustering.
}
 \label{fig:w_multipole}
\end{figure*}

\begin{figure*}
\begin{subfigure}[t]{\columnwidth}
\centering
 \includegraphics[width=\columnwidth]{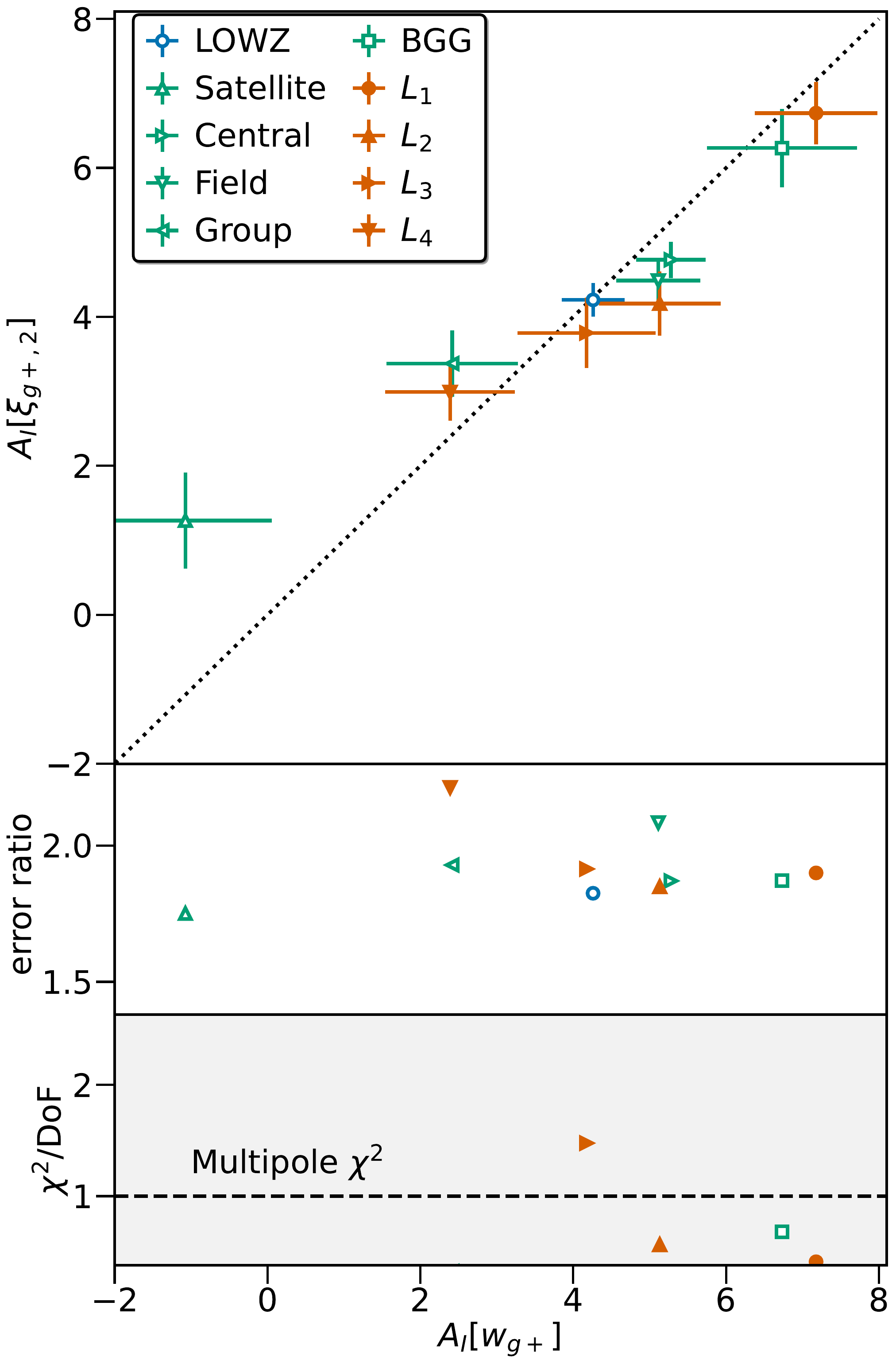}
\caption{}
\label{fig:AI_multipole_wgp}
\end{subfigure}
\begin{subfigure}[t]{\columnwidth}
\includegraphics[width=\columnwidth]{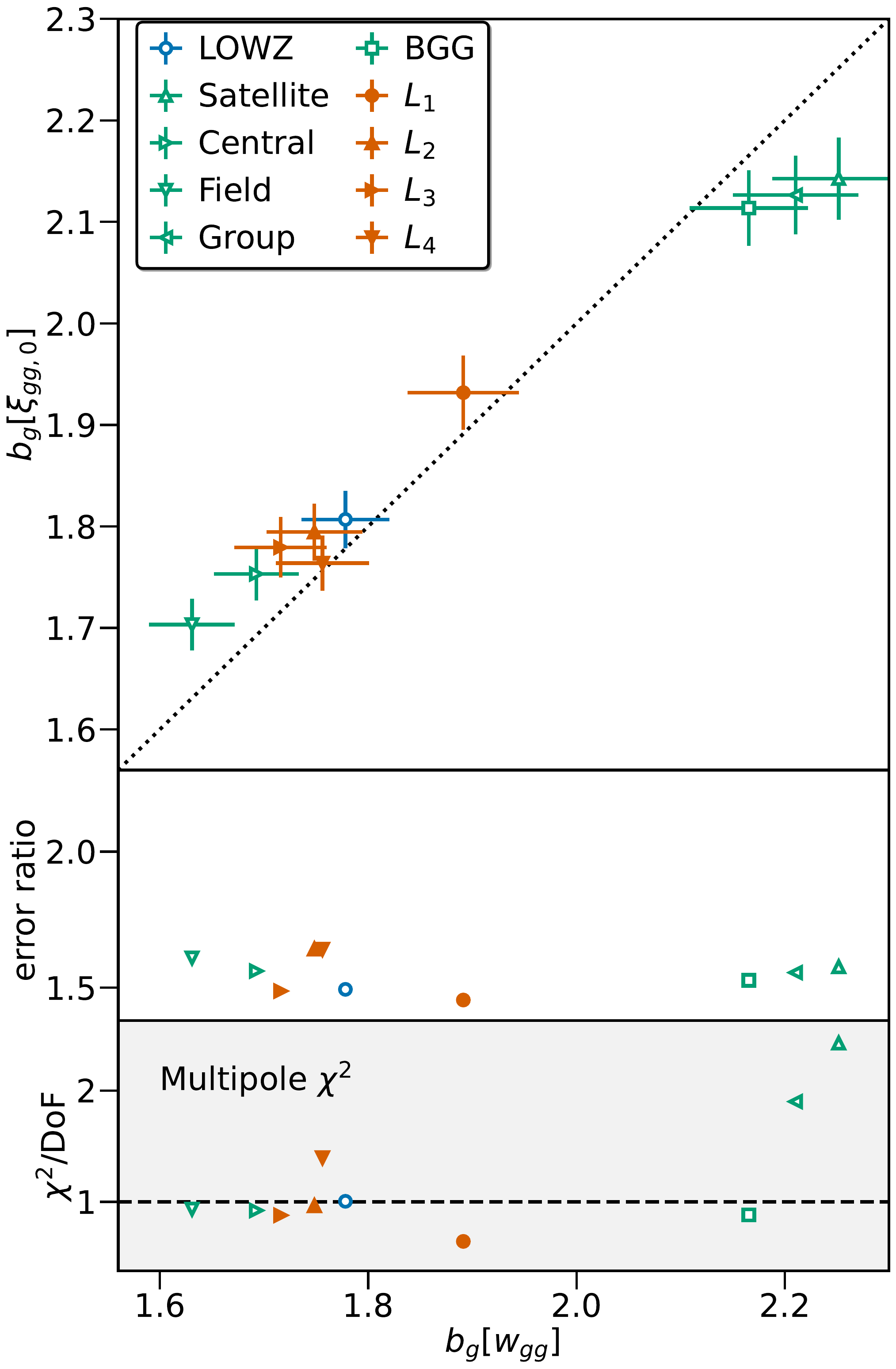}
\caption{}
\label{fig:bg_multipole_wgg}
\end{subfigure}
\caption{ \emph{Upper Panel}: Comparison of the model parameters, intrinsic alignment amplitude $A_I$ and galaxy bias, $b_g$ obtained from fitting the 
 wedged multipole moments of correlation functions (y-axis) and the projected correlation functions (x-axis) for subsamples of the LOWZ sample.
The parameters obtained using the two estimators are consistent with each other, indicating that the model is working consistently for both the estimators. Note that the parameters from the two 
estimators are not expected to match exactly or be tightly correlated since the two estimators are sensitive to different modes in the $(r,\mu)$ or equivalently the $(r_\mathrm{p}, \Pi$) plane.
\emph{Middle Panel}: The ratio of parameter uncertainties obtained from fits to the wedged multipoles versus those from the projected correlation functions. 
The multipole moments results in a better signal-to-noise ratio on the parameter estimates, with a factor of $1.5-2$ improvement across all 
 the subsamples.
The \emph{bottom panel} shows the separate $\chi^2$ contributions from $\widetilde{\xi}_{g+}$ and $\widetilde{\xi}_{gg}$  for the best fitting parameter values divided by the degrees of freedom (DoF $=16$), which is the number of $r$ or $r_\mathrm{p}$ bins fitted minus one (number of model parameters). 
}
\label{fig:model_comp}
\end{figure*}

\begin{figure}
\centering
\includegraphics[width=\columnwidth]{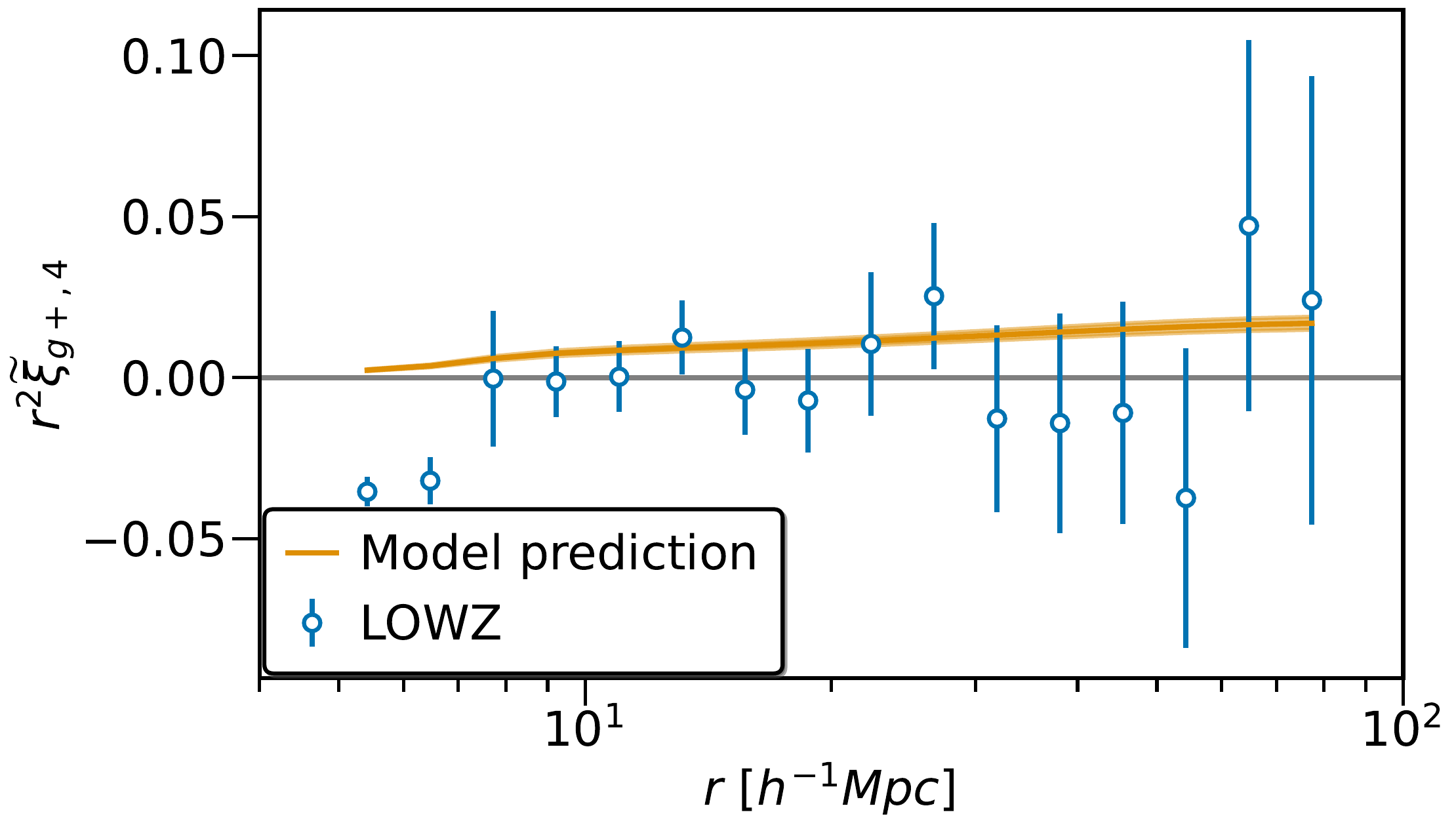}
\caption{Hexadecapole (wedge) of the galaxy-shape cross-correlation function, \xigp. The open markers with errorbars show the measurements from the full 
LOWZ sample, while the \refresponse{solid yellow line shows the NLA model prediction using the best-fitting parameters from fitting the quadrupole. The band around the model prediction shows $\pm2$ jackknife standard deviations of the model predictions.} The hexadecapole measurements are noisy compared to the predicted signal, and since our model also does not include the full non-linear RSD effects, we do not include the hexadecapole when fitting 
the model to the measurements.
 }
\label{fig:IA_hex}
\end{figure}

We begin by showing the galaxy clustering and galaxy-shape cross correlation measurements for the full LOWZ sample in the two dimensional space of $(r_\mathrm{p},\Pi)$ in Figure~\ref{fig:corr_2d}. 
The solid contours in the figure show the measurements from the data, while the solid lines show the best-fitting model (to \wgg\ and \wgp, as discussed later in this section). On large scales, the model is consistent with the data. However, on small scales ($r_\mathrm{p}\lesssim5\mpch$, marked by vertical dotted lines) there are significant deviations, primarily due to the Fingers-of-God effect. Since the model is not expected to explain the data well on these scales ($r_\mathrm{p}\lesssim5\mpch$), we exclude these scales when fitting the model and also when computing the wedged multipole moments. 
        		
In Figure~\ref{fig:w_multipole}, we show the measurements of both the projected correlation functions ($w_{gg}$, $w_{g+}$) and the lowest order nonzero wedged multipoles (monopole for $\widetilde{\xi}_{gg}$ and quadrupole for $\widetilde{\xi}_{g+}$). \refresponse{We constructed wedged multipoles with a $r_\mathrm{p}>5\mpch$ cut.  Then,  when fitting the models, we applied the cuts $r_\mathrm{p}>5~\mpch$ for projected correlation functions and $r>5~\mpch$ for the wedged multipoles.} Our simple model with only a linear galaxy bias and IA amplitude as free parameters fits the measurements quite well on these scales. It
is important to note that the qualities of these fits does not confirm that the model is complete on those scales, as there are nonzero clustering and IA contributions in higher multipoles \refresponse{than the monopole and quadrupole shown here.  For even higher multipoles}, the model may struggle to fit $\widetilde{\xi}_{gg}$ down to these scales. Modeling clustering multipoles is a well-explored problem in 
the literature and we refer the interested reader to publications such as \cite{SDSS2017,Vlah2019} and references therein for more information. 
        		
Our primary goal in this work is to show that it is possible to extract more information from the IA and clustering multipoles than from the projected estimators commonly used for IA measurements and model constraints. Figure~\ref{fig:w_multipole} shows this clearly as the multipole moments have a higher signal-to-noise ratio than the projected 
correlation functions. 
Furthermore, for the parameters of interest in IA modeling (in this case $A_I$), the constraints are improved by almost a factor of $\sim 1.8-2.2$ as shown by the $A_I$ values and uncertainties shown on this plot. 
The quality of the model fits and the parameter values are very similar for both the multipoles and the projected correlation functions, showing that it is better to use the lowest order multipoles instead of projected correlation functions even when the galaxy velocities are not fully modeled. 
        		
In Fig.~\ref{fig:model_comp}, we show the comparison of the IA model parameters obtained by splitting the LOWZ sample into various subsamples as defined in \cite{2017MNRAS.464.2120S}, encompassing different physical environments and different ranges of intrinsic luminosity.    In particular, we divide galaxies based on environment into satellite, central, field, group, and brightest group galaxies (BGGs); and into luminosity bins labeled $L_1$--$L_4$ based on intrinsic luminosities, with lower numbers corresponding to brighter samples.  By doing so, we can check the model quality for subsamples and also whether the improved constraints on IA model parameters persists across these samples. \refresponse{In addition, the $\chi^2$ values for all fits are shown in Table~\ref{tab:chi2}.}
\begin{table*}
	\centering
	
	 \begin{tabular}{|c c c c c c c|} 
	 \hline
		 Sample & $\chi^2_{\xi_{g+}}$ & $\chi^2_{\xi_{gg}}$ & $\chi^2_{w_{g+}}$ & $\chi^2_{w_{gg}}$ & $\Delta b$ & $\Delta A_I$\\ 
 \hline 
LOWZ  & 3.8  & 16.1  & 12.4  & 9.7  & $ -0.03 \pm 0.03 $  & $ 0.0 \pm 0.4 $\\  
 \hline 
Central  & 4.6  & 14.8  & 8.7  & 7.9  & $ -0.06 \pm 0.03 $  & $ 0.5 \pm 0.4 $\\  
 \hline 
Satellite  & 3.8  & 38.8  & 7.1  & 10.4  & $ 0.11 \pm 0.04 $  & $ -2.3 \pm 0.9 $\\  
 \hline 
Field  & 4.2  & 15.0  & 10.0  & 5.2  & $ -0.07 \pm 0.03 $  & $ 0.6 \pm 0.5 $\\  
 \hline 
Group  & 5.2  & 30.4  & 14.1  & 11.4  & $ 0.08 \pm 0.04 $  & $ -1.0 \pm 0.7 $\\  
 \hline 
BGG  & 10.9  & 14.1  & 19.5  & 11.6  & $ 0.05 \pm 0.04 $  & $ 0.5 \pm 0.8 $\\  
 \hline 
$L_1$  & 6.6  & 10.3  & 11.7  & 11.8  & $ -0.04 \pm 0.03 $  & $ 0.4 \pm 0.6 $\\  
 \hline 
$L_2$  & 9.0  & 15.4  & 17.5  & 10.2  & $ -0.05 \pm 0.03 $  & $ 1.0 \pm 0.6 $\\  
 \hline 
$L_3$  & 23.6  & 14.1  & 11.5  & 9.4  & $ -0.06 \pm 0.03 $  & $ 0.4 \pm 0.8 $\\  
 \hline 
$L_4$  & 4.1  & 22.3  & 11.6  & 5.3  & $ -0.01 \pm 0.03 $  & $ -0.6 \pm 0.8 $\\  
 \hline 
  	\end{tabular}
	\caption{\refresponse{
	$\chi^2$ values for fits to multipoles and projected correlation functions for various LOWZ subsamples. 
 All fits have 16 degrees of freedom. $\Delta b$ and $\Delta A_I$ are the differences in the parameters obtained from fitting projected correlations versus the multipoles ($b_{w_{gg}}-b_{\xi_{gg}}$ and $A_{I,{w_{g+}}}-A_{I,\xi_{g+}}$).}}
	\label{tab:chi2}
	\end{table*}
        
Both the galaxy bias and IA amplitudes obtained from the two estimators are 
\refresponse{statistically} consistent in value \refresponse{as shown in this figure and from the differences (with uncertainties) shown in Table~\ref{tab:chi2}.  Note that the parameters from the two 
estimators are not expected to match exactly or be tightly correlated since the two estimators are sensitive to different modes in the $(r,\mu)$ or $(r_\mathrm{p}, \Pi$) plane.  The only noticeable tension is in the galaxy bias for the satellite subsample, for which the $\chi^2$ value hints at a poor model fit as discussed below. Considering the uncertainties rather than the best-fitting values, we find that} the wedged multipoles consistently produc\refresponse{e} higher signal-to-noise measurements \refresponse{than the projected correlation functions}, resulting in factors of $\sim 1.5-1.8$ improvement in the galaxy bias constraints and factors of $\sim 1.8-2$ improvement in $A_I$. 

Since the ability of the model to explain the multipoles at small scales is a
concern due to the non-linear effects of redshift space distortions, we show the $\chi^2$ of the best-fitting models in the lower panel of Fig.~\ref{fig:model_comp} \refresponse{and in Table~\ref{tab:chi2}}. For galaxy clustering, the reduced $\chi^2$ is consistent with the expected value of 1, except 
for the samples dominated by satellite galaxies, namely the `group' sample containing only galaxies within the groups of 2 or more and the `satellite' sample which only selects satellites from the groups. \refresponse{For both of these, the given $\chi^2$ and number of degrees of freedom result in a $p$-value below 0.02, suggesting the tension reflects a real failure in our model, whereas for all other samples the $p$-value exceeds a standard threshold of 0.05, indicating the model is acceptable. The satellite and group} samples should have the largest
effects from non-linear RSD, \refresponse{which we do not include in our model,} and hence as expected the model cannot fully describe the data for these samples. For the case of galaxy shape cross correlations, we obtain $\chi^2$ values that are \refresponse{typically} $\lesssim 1$, indicating that  
the model works well and our jackknife covariance may have overestimated the errors somewhat. \refresponse{The worst case shown is the $L_3$ sample, for which the $\chi^2$ and number of degrees of freedom gives a $p$-value of 0.08 (meaning we cannot rule out the model based on goodness of fit using a standard threshold of 0.05).}
        
In Fig.~\ref{fig:IA_hex} we show the hexadecapole (wedge) measurements for $\widetilde{\xi}_{g+}$ using the full LOWZ sample. The hexadecapole contribution to $\widetilde{\xi}_{g+}$ arises purely because of the redshift space 
distortions in the galaxy positions. As shown \refresponse{by the close match between solid and dotted lines} in Fig.~\ref{fig:xigp_2d}, these RSD effects are relatively small and hence the hexadecapole signal is expected to be 
small and noisy. Fig.~\ref{fig:IA_hex} confirms this expectation, as the measurement is consistent with 
zero on most scales. We also show the model predictions for comparison and data is consistent with the model on most scales, except for very small scales where 
our model is not expected to capture the non-linear RSD effects very well. Because of the noise in the measurements and the incompleteness of our model, we do
not include the hexadecapole when fitting the model to the measurements.

\subsection{Results from the simulation}

To gain deeper insight into these results, we also performed the measurements using a state-of-the-art hydrodynamical simulation, the Illustris-TNG (for more details see Section~\ref{sec:data}).  Use of a simulation enables us to explore effects that cannot be separated in the real data, such as the impact of real space vs.\ redshift space. When fitting the models to the simulated data, 
we applied the cuts $r_\mathrm{p}>2\mpch$ when computing multipole wedges and also a further cut of $r>4\mpch$ when fitting the models. 
Given the smaller size of the simulation box, the lower $r_\mathrm{p}$ cut is necessary to obtain sufficient signal to noise in the parameter constraints.  Our tests with larger $r_\mathrm{p}$ cuts suggest our analysis is not biased (within the noise of the measurements).  
        
\begin{figure*}
\centering
\begin{subfigure}[t]{0.45\textwidth}
\centering
\includegraphics[width=1.\linewidth]{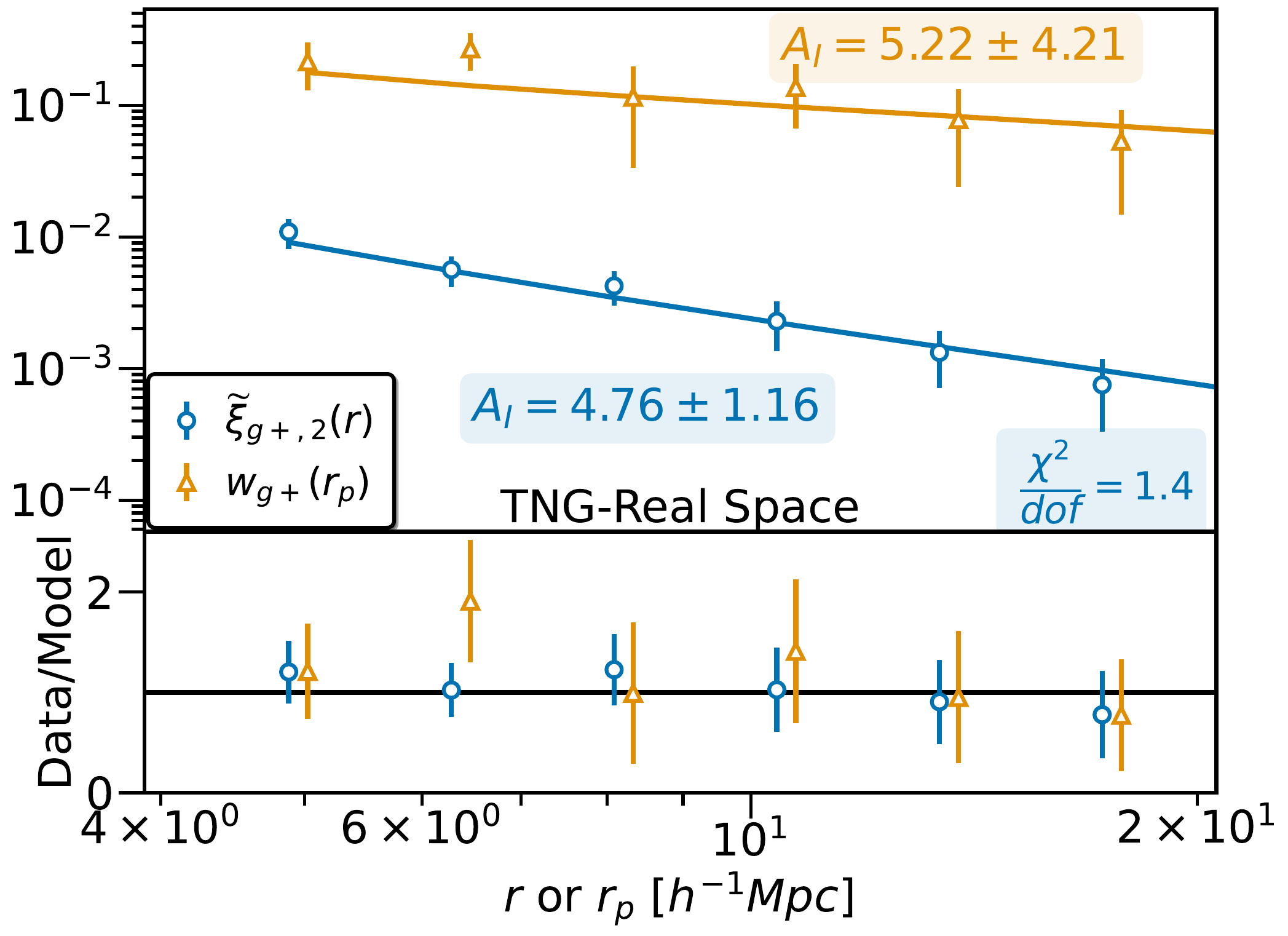} 
\caption{} \label{fig:timing1}
\end{subfigure}
\begin{subfigure}[t]{0.45\textwidth}
\centering
\includegraphics[width=1.\linewidth]{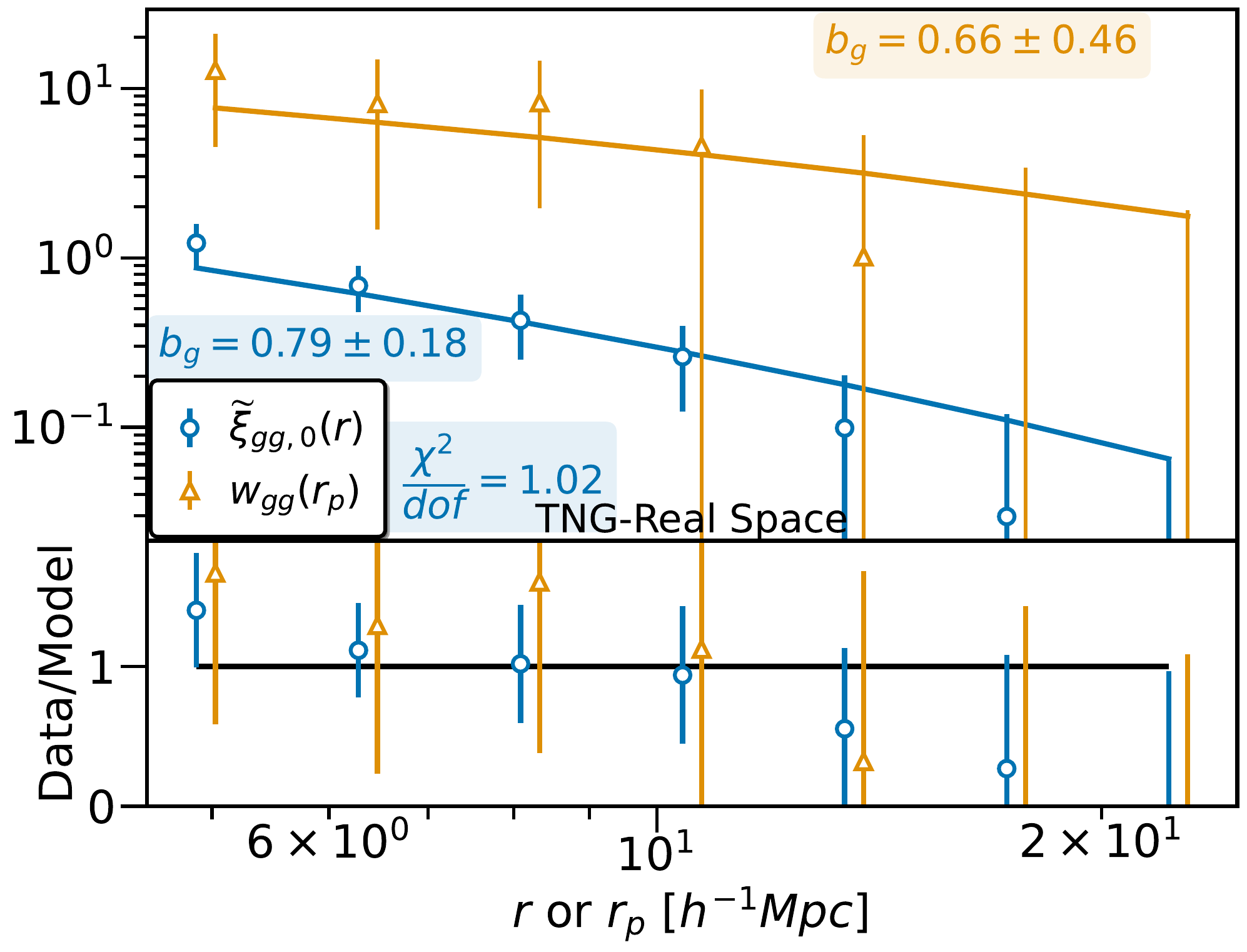} 
 \caption{} \label{fig:timing2}
\end{subfigure}
        
\vspace{1cm}
 \begin{subfigure}[t]{0.45\textwidth}
 \centering
 \includegraphics[width=1.\linewidth]{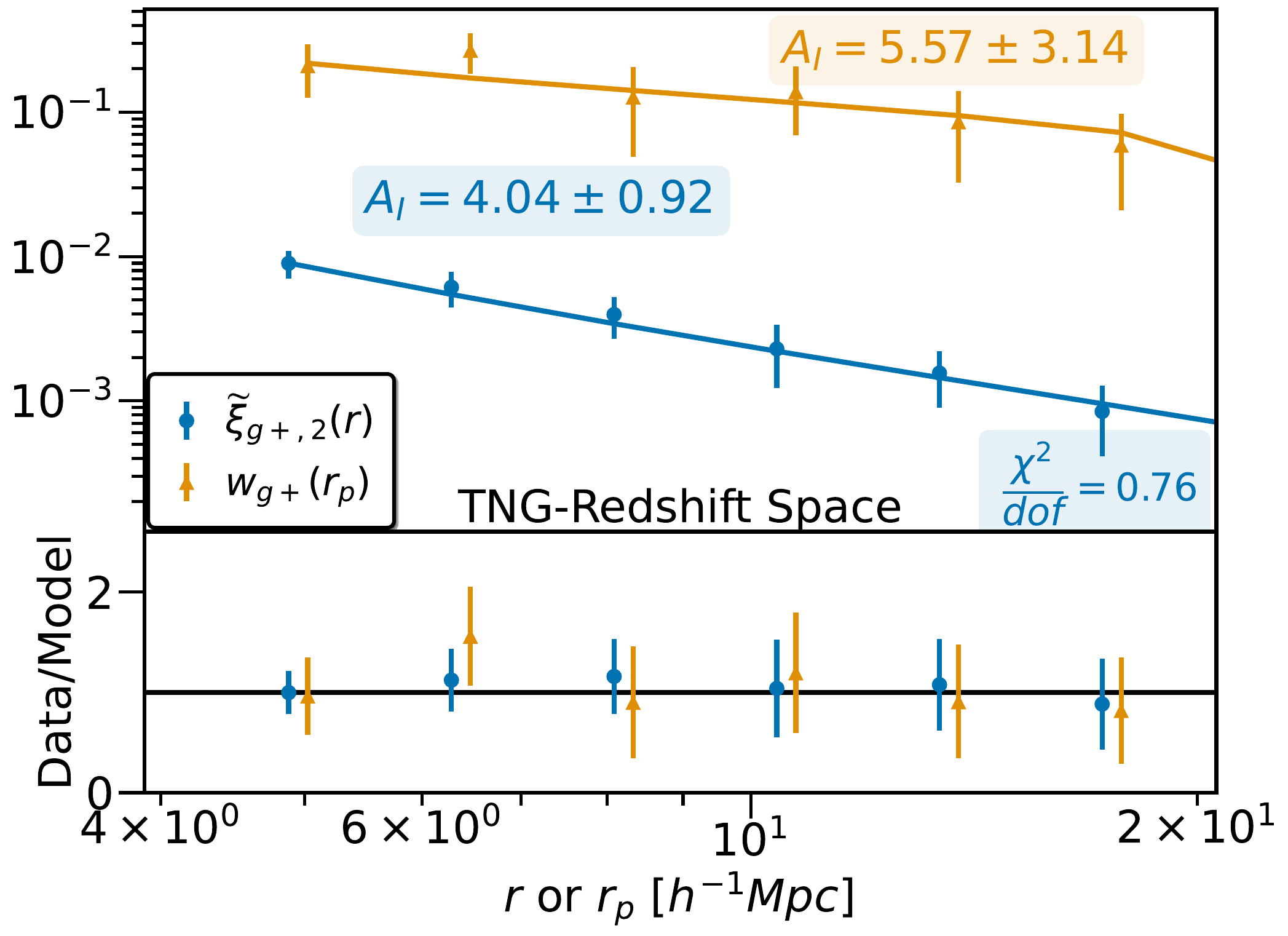} 
\caption{} \label{fig:timing3}
\end{subfigure}
\begin{subfigure}[t]{0.45\textwidth}
            
\centering
\includegraphics[width=1.\linewidth]{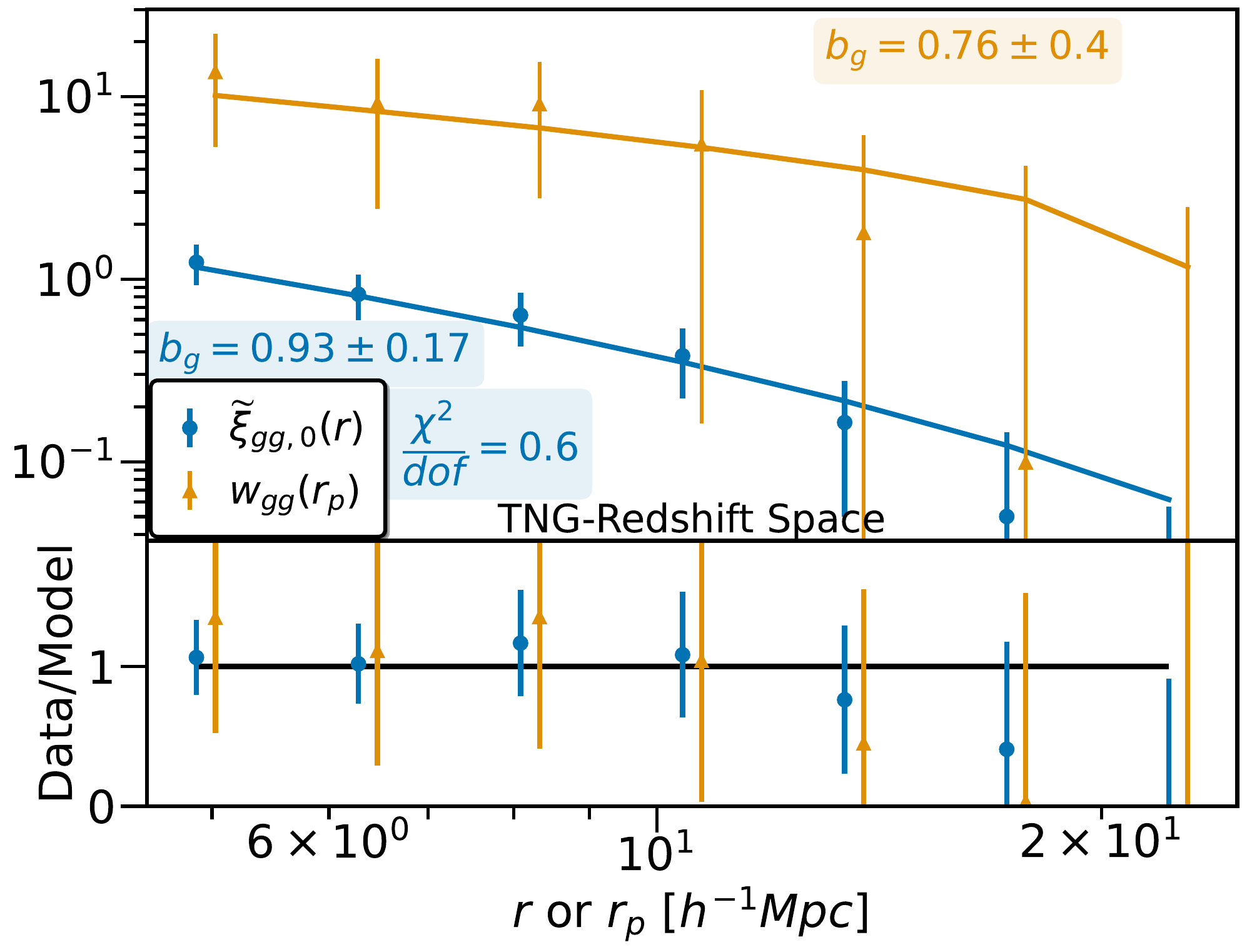} 
\caption{} \label{fig:timing4}
\end{subfigure}
         
\caption{ $\widetilde{\xi}_{g+}$ wedged quadrupole (1st column) and $\widetilde{\xi}_{gg}$ wedged monopole (2nd column) measurements from the elliptical galaxy sample selected in the IllustrisTNG-100 $z=0$ snapshot, along with the corresponding projected correlation functions. Here we show the wedged multipoles with $r_\mathrm{p}>2\mpch$ cuts and we use $r_\mathrm{min}>4\mpch$ for the fits. The first row shows measurements made in real space, whereas the second row is in redshift space. The  points show the 
measurements from the simulation with jackknife errorbars, while the solid lines show the best fitting model. The lower panel in all four figures shows the ratio of the measurements to the model. 
The model is in good agreement with the data on all scales. 
There is no significant difference between the \textit{real space} and the \textit{redshift space} measurements for the quantities shown here. 
\refresponse{Note that in the top row, the model does not include redshift-space distortions (since the measurements are made in configuration space), while in the bottom row, the model uses the linear (Kaiser) redshift-space distortion model.}}  \label{fig:tng}
\end{figure*}

In Figure~\ref{fig:tng}, we show the measured multipoles and their fits both in \textit{real space} and in \textit{redshift space}. The figure reiterates our findings in observed data:  that the parameter constraints are improved when using multipoles while the quality of the fits is also good. The $\chi^2$  values per degree of freedom  for each of the fits are within 40\% of the expected value of 1 \refresponse{given the 10 degrees of freedom}. For measurements done in real space, the constraints on $A_I$ are tighter by a factor 4 than those using projected correlation functions, with  reduced $\chi^2=0.79$ \refresponse{with 10 degrees of freedom}, indicating a  well-fit model. \refresponse{While the models for the galaxy clustering ($gg$) correlations tend to be systematically above the data on large scales in Fig.~\ref{fig:tng}, this can likely be explained by the large off-diagonal terms in the covariances (see Fig.~\ref{fig:correlationmatrices}) rather than reflecting a failure in the model.  The correlations mean that in general the fit will prefer that the data always be systematically above or below the model on large scales, ruling out solutions that have the model above some points and below others.}

To further investigate the effects of galaxy velocities, we performed the same measurements but in redshift space (row~2 of Fig.~\ref{fig:tng}).  The trends for these fits are similar: we get an improvement of almost a factor of 3 in the constraint on $A_I$, while the quality of 
fits is not degraded because of un-modeled non-linear velocity effects.  All in all, these results suggest that using multipoles to measure the IA signal improves the constraints on the parameters of interest by a substantial amount, and measurements done in \textit{real space} and \textit{redshift space} show consistent results.
  
\section{Conclusions}\label{sec:conclusions}


In this paper, we have demonstrated the performance of a multipole-based estimator for the intrinsic alignment signal, to be employed in direct measurements of alignments using galaxy samples with shapes and spectroscopic redshifts. 
The goal of using this estimator is to enable a more optimal extraction of information about intrinsic alignment models from the measurements, effectively increasing the statistical constraining power of current and future datasets.  In practice, it is only necessary to use the lowest nonzero multipole (i.e., the quadrupole) for the intrinsic alignment signal in order to unlock the benefit of this new estimator.

By applying this estimator to both simulated and real data, we have identified substantial gains in constraining power.  For example, when interpreting the measurements using the nonlinear alignment (NLA) model \citep{Bridle2007}, we found a factor of $\sim2.3$ increase in statistical precision of the alignment amplitude in simulations, and a doubling of the statistical precision in real data.  This is equivalent to multiplying the survey area by a factor of 4, given that the constraints are shape noise dominated.

It is worth keeping in mind some potential future improvements in modeling that will be needed to fully exploit the improved constraining power in this multipole-based estimator. First, making use of the improved statistical power in the measurements with multipoles requires that one be able to produce a theoretical prediction in the $(r_\mathrm{p}, \Pi)$ plane so as to calculate multipoles and produce quantities that can be compared with this estimator. While this is clearly already the case for the NLA model, it may not be as straightforward for other intrinsic alignment methods of interest in the field. 
 However, in most models, the angular dependence is separable from the scale dependence, i.e., the $\mu^2$-dependence is separable from the $k$-dependence, and can likely be modeled with a limited number of multipoles in some cases; see, for example, eq.~2.63 in \cite{Bakx2023}.  
Of course, this limitation also exists for projected correlation functions computed with finite line of sight integration length \citep{Baldauf2010, 2019MNRAS.482..785S}. 

Another potential issue in \refresponse{the} future is that the increased signal-to-noise in the measurements will necessitate an improved model for interpreting those measurements.  Our results suggest that the current model with Halofit \refresponse{\citep{Takahashi2012}} and the Kaiser correction is sufficient to model the 
lowest order multipole wedges (monopole for galaxy auto-correlations and quadrupole for galaxy-shape correlations) using $r_\mathrm{p}>5\mpch$. $r_\mathrm{p}\sim5\mpch$ is also the scale to which projected correlation functions can be modeled using the same model \citep{Singh2015}.  However, this new estimator may require an improved theoretical model with other samples or to push to smaller physical scales. 

Finally, while direct IA measurements can be made using high-quality photometric redshifts, additional development would be needed to ascertain how to use a multipole-based estimator in such a case and whether it adds any advantage. 
\refresponse{P}hotometric redshift uncertainties modify the expected shape of the signal in the $(r_p,\Pi)$ plane, smearing out the correlation function along both the $r_p$ and $ \Pi$ directions, qualitatively similar to RSD effects but with opposite sign and an order of magnitude larger strength (typically, $\Delta z\sim10^{-3}$ for RSD and $\Delta z\gtrsim 0.01$ for high-quality photometric redshifts).
This would require additional modeling and using higher order multipole moments to optimally extract the signal.

In the near future, we anticipate that intrinsic alignment constraints with existing datasets could be productively re-evaluated using this estimator.  The resulting improvements in our understanding of intrinsic alignments and their scaling with luminosity, colour, and more \citep[e.g.,][]{Samuroff2022} has great potential to improve the systematics mitigation process for cosmological weak lensing measurements with ongoing and upcoming surveys. 

\section*{Data Availability Statement}

The simulation catalog data is available at \url{https://github.com/McWilliamsCenter/gal_decomp_paper}, while the SDSS-III BOSS LOWZ catalog is available at \url{https://cosmo.vera.psc.edu/SDSS_shape_catalog/}.
The software used to calculate the correlation functions in this work is available at \url{https://github.com/sukhdeep2/corr_pc}, and the analysis codes will be made available upon reasonable request to the authors.

\section*{Acknowledgments}

We thank Elisa Chisari and Jonathan Blazek for useful discussions related to this work.
\refresponse{We thank the anonymous referee for feedback that has significantly improved the presentation of this work.}

SS is supported by McWilliams postdoctoral fellowship at CMU.
AS was supported by a CMU Summer Undergraduate Research Fellowship (SURF). RM and YJ were supported in part by Department of Energy grant
DE-SC0010118 and in part by a grant from the Simons Foundation (Simons
Investigator in Astrophysics, Award ID 620789). 

\bibliographystyle{mnras}
  \bibliography{IA,refs}

\begin{thebibliography}{}
\makeatletter
\relax
\def\mn@urlcharsother{\let\do\@makeother \do\$\do\&\do\#\do\^\do\_\do\%\do\~}
\def\mn@doi{\begingroup\mn@urlcharsother \@ifnextchar [ {\mn@doi@}
  {\mn@doi@[]}}
\def\mn@doi@[#1]#2{\def\@tempa{#1}\ifx\@tempa\@empty \href
  {http://dx.doi.org/#2} {doi:#2}\else \href {http://dx.doi.org/#2} {#1}\fi
  \endgroup}
\def\mn@eprint#1#2{\mn@eprint@#1:#2::\@nil}
\def\mn@eprint@arXiv#1{\href {http://arxiv.org/abs/#1} {{\tt arXiv:#1}}}
\def\mn@eprint@dblp#1{\href {http://dblp.uni-trier.de/rec/bibtex/#1.xml}
  {dblp:#1}}
\def\mn@eprint@#1:#2:#3:#4\@nil{\def\@tempa {#1}\def\@tempb {#2}\def\@tempc
  {#3}\ifx \@tempc \@empty \let \@tempc \@tempb \let \@tempb \@tempa \fi \ifx
  \@tempb \@empty \def\@tempb {arXiv}\fi \@ifundefined
  {mn@eprint@\@tempb}{\@tempb:\@tempc}{\expandafter \expandafter \csname
  mn@eprint@\@tempb\endcsname \expandafter{\@tempc}}}

\bibitem[\protect\citeauthoryear{{Abazajian} et~al.,}{{Abazajian}
  et~al.}{2009}]{2009ApJS..182..543A}
{Abazajian} K.~N.,  et~al., 2009, \mn@doi [\apjs]
  {10.1088/0067-0049/182/2/543}, \href
  {http://adsabs.harvard.edu/abs/2009ApJS..182..543A} {182, 543}

\bibitem[\protect\citeauthoryear{{Ahn} et~al.,}{{Ahn} et~al.}{2012}]{Ahn:2012}
{Ahn} C.~P.,  et~al., 2012, \mn@doi [\apjs] {10.1088/0067-0049/203/2/21}, \href
  {http://adsabs.harvard.edu/abs/2012ApJS..203...21A} {203, 21}

\bibitem[\protect\citeauthoryear{{Aihara} et~al.,}{{Aihara}
  et~al.}{2011}]{2011ApJS..193...29A}
{Aihara} H.,  et~al., 2011, \mn@doi [\apjs] {10.1088/0067-0049/193/2/29}, \href
  {http://adsabs.harvard.edu/abs/2011ApJS..193...29A} {193, 29}

\bibitem[\protect\citeauthoryear{{Akeson} et~al.,}{{Akeson}
  et~al.}{2019}]{akeson19}
{Akeson} R.,  et~al., 2019, preprint, \href
  {https://ui.adsabs.harvard.edu/abs/2019arXiv190205569A} {} (\mn@eprint
  {arXiv} {1902.05569})

\bibitem[\protect\citeauthoryear{{Alam} et~al.,}{{Alam}
  et~al.}{2017}]{SDSS2017}
{Alam} S.,  et~al., 2017, \mn@doi [\mnras] {10.1093/mnras/stx721}, \href
  {https://ui.adsabs.harvard.edu/abs/2017MNRAS.470.2617A} {470, 2617}

\bibitem[\protect\citeauthoryear{{Amon} et~al.,}{{Amon}
  et~al.}{2022}]{Amon2022}
{Amon} A.,  et~al., 2022, \mn@doi [\prd] {10.1103/PhysRevD.105.023514}, \href
  {https://ui.adsabs.harvard.edu/abs/2022PhRvD.105b3514A} {105, 023514}

\bibitem[\protect\citeauthoryear{{Bakx}, {Kurita}, {Chisari}, {Vlah}  \&
  {Schmidt}}{{Bakx} et~al.}{2023}]{Bakx2023}
{Bakx} T.,  {Kurita} T.,  {Chisari} N.~E.,  {Vlah} Z.,   {Schmidt} F.,  2023,
  \mn@doi [arXiv e-prints] {10.48550/arXiv.2303.15565}, \href
  {https://ui.adsabs.harvard.edu/abs/2023arXiv230315565B} {p. arXiv:2303.15565}

\bibitem[\protect\citeauthoryear{{Baldauf}, {Smith}, {Seljak}  \&
  {Mandelbaum}}{{Baldauf} et~al.}{2010}]{Baldauf2010}
{Baldauf} T.,  {Smith} R.~E.,  {Seljak} U.,   {Mandelbaum} R.,  2010, \mn@doi
  [\prd] {10.1103/PhysRevD.81.063531}, \href
  {http://adsabs.harvard.edu/abs/2010PhRvD..81f3531B} {81, 063531}

\bibitem[\protect\citeauthoryear{{Blazek}, {Vlah}  \& {Seljak}}{{Blazek}
  et~al.}{2015}]{Blazek2015}
{Blazek} J.,  {Vlah} Z.,   {Seljak} U.,  2015, \mn@doi [\jcap]
  {10.1088/1475-7516/2015/08/015}, \href
  {https://ui.adsabs.harvard.edu/abs/2015JCAP...08..015B} {2015, 015}

\bibitem[\protect\citeauthoryear{{Bolton} et~al.,}{{Bolton}
  et~al.}{2012}]{Bolton:2012}
{Bolton} A.~S.,  et~al., 2012, \mn@doi [\aj] {10.1088/0004-6256/144/5/144},
  \href {http://adsabs.harvard.edu/abs/2012AJ....144..144B} {144, 144}

\bibitem[\protect\citeauthoryear{{Bridle} \& {King}}{{Bridle} \&
  {King}}{2007}]{Bridle2007}
{Bridle} S.,  {King} L.,  2007, \mn@doi [New Journal of Physics]
  {10.1088/1367-2630/9/12/444}, \href
  {http://adsabs.harvard.edu/abs/2007NJPh....9..444B} {9, 444}

\bibitem[\protect\citeauthoryear{{Catelan}, {Kamionkowski}  \&
  {Blandford}}{{Catelan} et~al.}{2001}]{Catelan2001}
{Catelan} P.,  {Kamionkowski} M.,   {Blandford} R.~D.,  2001, \mn@doi [\mnras]
  {10.1046/j.1365-8711.2001.04105.x}, \href
  {http://adsabs.harvard.edu/abs/2001MNRAS.320L...7C} {320, L7}

\bibitem[\protect\citeauthoryear{{Dalal} et~al.,}{{Dalal}
  et~al.}{2023}]{Dalal2023}
{Dalal} R.,  et~al., 2023, preprint, \href
  {https://ui.adsabs.harvard.edu/abs/2023arXiv230400701D} {} (\mn@eprint
  {arXiv} {2304.00701})

\bibitem[\protect\citeauthoryear{{Davis}, {Efstathiou}, {Frenk}  \&
  {White}}{{Davis} et~al.}{1985}]{fof}
{Davis} M.,  {Efstathiou} G.,  {Frenk} C.~S.,   {White} S.~D.~M.,  1985,
  \mn@doi [\apj] {10.1086/163168}, \href
  {https://ui.adsabs.harvard.edu/abs/1985ApJ...292..371D} {292, 371}

\bibitem[\protect\citeauthoryear{{Dawson} et~al.,}{{Dawson}
  et~al.}{2013}]{Dawson:2013}
{Dawson} K.~S.,  et~al., 2013, \mn@doi [\aj] {10.1088/0004-6256/145/1/10},
  \href {http://adsabs.harvard.edu/abs/2013AJ....145...10D} {145, 10}

\bibitem[\protect\citeauthoryear{{Eisenstein} et~al.,}{{Eisenstein}
  et~al.}{2001}]{2001AJ....122.2267E}
{Eisenstein} D.~J.,  et~al., 2001, \aj, \href
  {http://adsabs.harvard.edu/cgi-bin/nph-bib_query?bibcode=2001AJ....122.2267E&amp;db_key=AST}
  {122, 2267}

\bibitem[\protect\citeauthoryear{{Fortuna}, {Hoekstra}, {Joachimi}, {Johnston},
  {Chisari}, {Georgiou}  \& {Mahony}}{{Fortuna} et~al.}{2021a}]{Fortuna2021hm}
{Fortuna} M.~C.,  {Hoekstra} H.,  {Joachimi} B.,  {Johnston} H.,  {Chisari}
  N.~E.,  {Georgiou} C.,   {Mahony} C.,  2021a, \mn@doi [\mnras]
  {10.1093/mnras/staa3802}, \href
  {https://ui.adsabs.harvard.edu/abs/2021MNRAS.501.2983F} {501, 2983}

\bibitem[\protect\citeauthoryear{{Fortuna} et~al.,}{{Fortuna}
  et~al.}{2021b}]{Fortuna2021}
{Fortuna} M.~C.,  et~al., 2021b, \mn@doi [\aap] {10.1051/0004-6361/202140706},
  \href {https://ui.adsabs.harvard.edu/abs/2021A&A...654A..76F} {654, A76}

\bibitem[\protect\citeauthoryear{{Fukugita}, {Ichikawa}, {Gunn}, {Doi},
  {Shimasaku}  \& {Schneider}}{{Fukugita} et~al.}{1996}]{1996AJ....111.1748F}
{Fukugita} M.,  {Ichikawa} T.,  {Gunn} J.~E.,  {Doi} M.,  {Shimasaku} K.,
  {Schneider} D.~P.,  1996, \aj, \href
  {http://adsabs.harvard.edu/cgi-bin/nph-bib_query?bibcode=1996AJ....111.1748F&amp;db_key=AST}
  {111, 1748}

\bibitem[\protect\citeauthoryear{{Gunn} et~al.,}{{Gunn}
  et~al.}{1998}]{1998AJ....116.3040G}
{Gunn} J.~E.,  et~al., 1998, \aj, \href
  {http://adsabs.harvard.edu/cgi-bin/nph-bib_query?bibcode=1998AJ....116.3040G&db_key=AST}
  {116, 3040}

\bibitem[\protect\citeauthoryear{{Gunn} et~al.,}{{Gunn}
  et~al.}{2006}]{Gunn2006}
{Gunn} J.~E.,  et~al., 2006, \mn@doi [\aj] {10.1086/500975}, \href
  {http://adsabs.harvard.edu/abs/2006AJ....131.2332G} {131, 2332}

\bibitem[\protect\citeauthoryear{{Hartlap}, {Simon}  \& {Schneider}}{{Hartlap}
  et~al.}{2007}]{Hartlap2007}
{Hartlap} J.,  {Simon} P.,   {Schneider} P.,  2007, \mn@doi [\aap]
  {10.1051/0004-6361:20066170}, \href
  {http://adsabs.harvard.edu/abs/2007A%26A...464..399H} {464, 399}

\bibitem[\protect\citeauthoryear{{Hirata} \& {Seljak}}{{Hirata} \&
  {Seljak}}{2004}]{Hirata2004}
{Hirata} C.~M.,  {Seljak} U.,  2004, \mn@doi [\prd]
  {10.1103/PhysRevD.70.063526}, \href
  {http://adsabs.harvard.edu/abs/2004PhRvD..70f3526H} {70, 063526}

\bibitem[\protect\citeauthoryear{{Hirata}, {Mandelbaum}, {Ishak}, {Seljak},
  {Nichol}, {Pimbblet}, {Ross}  \& {Wake}}{{Hirata} et~al.}{2007}]{Hirata2007}
{Hirata} C.~M.,  {Mandelbaum} R.,  {Ishak} M.,  {Seljak} U.,  {Nichol} R.,
  {Pimbblet} K.~A.,  {Ross} N.~P.,   {Wake} D.,  2007, \mn@doi [\mnras]
  {10.1111/j.1365-2966.2007.12312.x}, \href
  {https://ui.adsabs.harvard.edu/abs/2007MNRAS.381.1197H} {381, 1197}

\bibitem[\protect\citeauthoryear{{Hogg}, {Finkbeiner}, {Schlegel}  \&
  {Gunn}}{{Hogg} et~al.}{2001}]{2001AJ....122.2129H}
{Hogg} D.~W.,  {Finkbeiner} D.~P.,  {Schlegel} D.~J.,   {Gunn} J.~E.,  2001,
  \aj, \href
  {http://adsabs.harvard.edu/cgi-bin/nph-bib_query?bibcode=2001AJ....122.2129H&amp;db_key=AST}
  {122, 2129}

\bibitem[\protect\citeauthoryear{{Ivezi{\'c}} et~al.,}{{Ivezi{\'c}}
  et~al.}{2004}]{2004AN....325..583I}
{Ivezi{\'c}} {\v Z}.,  et~al., 2004, Astronomische Nachrichten, \href
  {http://adsabs.harvard.edu/cgi-bin/nph-bib_query?bibcode=2004AN....325..583I&db_key=AST}
  {325, 583}

\bibitem[\protect\citeauthoryear{{Ivezi{\'c}} et~al.,}{{Ivezi{\'c}}
  et~al.}{2019}]{Ivezic2019}
{Ivezi{\'c}} {\v{Z}}.,  et~al., 2019, \mn@doi [\apj]
  {10.3847/1538-4357/ab042c}, \href
  {https://ui.adsabs.harvard.edu/abs/2019ApJ...873..111I} {873, 111}

\bibitem[\protect\citeauthoryear{{Jagvaral}, {Campbell}, {Mandelbaum}  \&
  {Rau}}{{Jagvaral} et~al.}{2022a}]{jagvaral-gal-decomp}
{Jagvaral} Y.,  {Campbell} D.,  {Mandelbaum} R.,   {Rau} M.~M.,  2022a, \mn@doi
  [\mnras] {10.1093/mnras/stab3104}, \href
  {https://ui.adsabs.harvard.edu/abs/2022MNRAS.509.1764J} {509, 1764}

\bibitem[\protect\citeauthoryear{{Jagvaral}, {Singh}  \&
  {Mandelbaum}}{{Jagvaral} et~al.}{2022b}]{jagvaral-bulge-disc-ia}
{Jagvaral} Y.,  {Singh} S.,   {Mandelbaum} R.,  2022b, \mn@doi [\mnras]
  {10.1093/mnras/stac1424}, \href
  {https://ui.adsabs.harvard.edu/abs/2022MNRAS.514.1021J} {514, 1021}

\bibitem[\protect\citeauthoryear{{Joachimi}, {Mandelbaum}, {Abdalla}  \&
  {Bridle}}{{Joachimi} et~al.}{2011}]{Joachimi2011}
{Joachimi} B.,  {Mandelbaum} R.,  {Abdalla} F.~B.,   {Bridle} S.~L.,  2011,
  \mn@doi [\aap] {10.1051/0004-6361/201015621}, \href
  {http://adsabs.harvard.edu/abs/2011A%26A...527A..26J} {527, A26}

\bibitem[\protect\citeauthoryear{{Joachimi} et~al.,}{{Joachimi}
  et~al.}{2015}]{Joachimi2015ia}
{Joachimi} B.,  et~al., 2015, \mn@doi [\ssr] {10.1007/s11214-015-0177-4}, \href
  {https://ui.adsabs.harvard.edu/abs/2015SSRv..193....1J} {193, 1}

\bibitem[\protect\citeauthoryear{{Johnston} et~al.,}{{Johnston}
  et~al.}{2019}]{Johnston2019}
{Johnston} H.,  et~al., 2019, \mn@doi [\aap] {10.1051/0004-6361/201834714},
  \href {https://ui.adsabs.harvard.edu/abs/2019A&A...624A..30J} {624, A30}

\bibitem[\protect\citeauthoryear{{Kaiser}}{{Kaiser}}{1987}]{Kaiser1987}
{Kaiser} N.,  1987, \mnras, \href
  {http://adsabs.harvard.edu/abs/1987MNRAS.227....1K} {227, 1}

\bibitem[\protect\citeauthoryear{{Kazin} et~al.,}{{Kazin}
  et~al.}{2013}]{Kazin2013}
{Kazin} E.~A.,  et~al., 2013, \mn@doi [\mnras] {10.1093/mnras/stt1261}, \href
  {https://ui.adsabs.harvard.edu/abs/2013MNRAS.435...64K} {435, 64}

\bibitem[\protect\citeauthoryear{{Kilbinger}}{{Kilbinger}}{2015}]{Kilbinger2015}
{Kilbinger} M.,  2015, \mn@doi [Reports on Progress in Physics]
  {10.1088/0034-4885/78/8/086901}, \href
  {http://adsabs.harvard.edu/abs/2015RPPh...78h6901K} {78, 086901}

\bibitem[\protect\citeauthoryear{{Kurita} \& {Takada}}{{Kurita} \&
  {Takada}}{2023}]{2023PhRvD.108h3533K}
{Kurita} T.,  {Takada} M.,  2023, \mn@doi [\prd] {10.1103/PhysRevD.108.083533},
  \href {https://ui.adsabs.harvard.edu/abs/2023PhRvD.108h3533K} {108, 083533}

\bibitem[\protect\citeauthoryear{{Kurita}, {Takada}, {Nishimichi}, {Takahashi},
  {Osato}  \& {Kobayashi}}{{Kurita} et~al.}{2021}]{Kurita2021}
{Kurita} T.,  {Takada} M.,  {Nishimichi} T.,  {Takahashi} R.,  {Osato} K.,
  {Kobayashi} Y.,  2021, \mn@doi [\mnras] {10.1093/mnras/staa3625}, \href
  {https://ui.adsabs.harvard.edu/abs/2021MNRAS.501..833K} {501, 833}

\bibitem[\protect\citeauthoryear{{Landy} \& {Szalay}}{{Landy} \&
  {Szalay}}{1993}]{landy}
{Landy} S.~D.,  {Szalay} A.~S.,  1993, \mn@doi [\apj] {10.1086/172900}, \href
  {https://ui.adsabs.harvard.edu/abs/1993ApJ...412...64L} {412, 64}

\bibitem[\protect\citeauthoryear{{Laureijs} et~al.,}{{Laureijs}
  et~al.}{2011}]{Laureijs2011}
{Laureijs} R.,  et~al., 2011, preprint, \href
  {https://ui.adsabs.harvard.edu/abs/2011arXiv1110.3193L} {} (\mn@eprint
  {arXiv} {1110.3193})

\bibitem[\protect\citeauthoryear{{Li}, {Singh}, {Yu}, {Feng}  \& {Seljak}}{{Li}
  et~al.}{2019}]{Li2019}
{Li} Y.,  {Singh} S.,  {Yu} B.,  {Feng} Y.,   {Seljak} U.,  2019, \mn@doi
  [\jcap] {10.1088/1475-7516/2019/01/016}, \href
  {https://ui.adsabs.harvard.edu/abs/2019JCAP...01..016L} {2019, 016}

\bibitem[\protect\citeauthoryear{{Li} et~al.,}{{Li} et~al.}{2023}]{Li2023}
{Li} X.,  et~al., 2023, preprint, \href
  {https://ui.adsabs.harvard.edu/abs/2023arXiv230400702L} {} (\mn@eprint
  {arXiv} {2304.00702})

\bibitem[\protect\citeauthoryear{{Lupton}, {Gunn}, {Ivezi{\'c}}, {Knapp}  \&
  {Kent}}{{Lupton} et~al.}{2001}]{2001ASPC..238..269L}
{Lupton} R.,  {Gunn} J.~E.,  {Ivezi{\'c}} Z.,  {Knapp} G.~R.,   {Kent} S.,
  2001, in {Harnden} F.~R. J.,  {Primini} F.~A.,   {Payne} H.~E.,  eds,
  Astronomical Society of the Pacific Conference Series Vol. 238, Astronomical
  Data Analysis Software and Systems X. p.~269 (\mn@eprint {arXiv}
  {astro-ph/0101420}), \mn@doi{10.48550/arXiv.astro-ph/0101420}

\bibitem[\protect\citeauthoryear{{Mandelbaum}, {Hirata}, {Ishak}, {Seljak}  \&
  {Brinkmann}}{{Mandelbaum} et~al.}{2006}]{Mandelbaum2006}
{Mandelbaum} R.,  {Hirata} C.~M.,  {Ishak} M.,  {Seljak} U.,   {Brinkmann} J.,
  2006, \mn@doi [\mnras] {10.1111/j.1365-2966.2005.09946.x}, \href
  {http://adsabs.harvard.edu/abs/2006MNRAS.367..611M} {367, 611}

\bibitem[\protect\citeauthoryear{{Mandelbaum} et~al.,}{{Mandelbaum}
  et~al.}{2011}]{mandelbaum-2011}
{Mandelbaum} R.,  et~al., 2011, \mn@doi [\mnras]
  {10.1111/j.1365-2966.2010.17485.x}, \href
  {https://ui.adsabs.harvard.edu/abs/2011MNRAS.410..844M} {410, 844}

\bibitem[\protect\citeauthoryear{{Mandelbaum}, {Slosar}, {Baldauf}, {Seljak},
  {Hirata}, {Nakajima}, {Reyes}  \& {Smith}}{{Mandelbaum}
  et~al.}{2013}]{Mandelbaum2013}
{Mandelbaum} R.,  {Slosar} A.,  {Baldauf} T.,  {Seljak} U.,  {Hirata} C.~M.,
  {Nakajima} R.,  {Reyes} R.,   {Smith} R.~E.,  2013, \mn@doi [\mnras]
  {10.1093/mnras/stt572}, \href
  {http://adsabs.harvard.edu/abs/2013MNRAS.432.1544M} {432, 1544}

\bibitem[\protect\citeauthoryear{{Mandelbaum} et~al.,}{{Mandelbaum}
  et~al.}{2018}]{Mandelbaum2018}
{Mandelbaum} R.,  et~al., 2018, \mn@doi [\mnras] {10.1093/mnras/sty2420}, \href
  {https://ui.adsabs.harvard.edu/abs/2018MNRAS.481.3170M} {481, 3170}

\bibitem[\protect\citeauthoryear{{Manera} et~al.,}{{Manera}
  et~al.}{2015}]{Manera2015}
{Manera} M.,  et~al., 2015, \mn@doi [\mnras] {10.1093/mnras/stu2465}, \href
  {http://adsabs.harvard.edu/abs/2015MNRAS.447..437M} {447, 437}

\bibitem[\protect\citeauthoryear{Marinacci et~al.}{Marinacci
  et~al.}{2018}]{Marinacci2017illustristng}
Marinacci F.,  et~al., 2018, \mn@doi [Mon. Not. Roy. Astron. Soc.]
  {10.1093/mnras/sty2206}, 480, 5113

\bibitem[\protect\citeauthoryear{{Naiman} et~al.,}{{Naiman}
  et~al.}{2018}]{Naiman2018illustristng}
{Naiman} J.~P.,  et~al., 2018, \mn@doi [\mnras] {10.1093/mnras/sty618}, \href
  {https://ui.adsabs.harvard.edu/abs/2018MNRAS.477.1206N} {477, 1206}

\bibitem[\protect\citeauthoryear{Nelson et~al.}{Nelson
  et~al.}{2018}]{tng-bimodal}
Nelson D.,  et~al., 2018, \mn@doi [Mon. Not. Roy. Astron. Soc.]
  {10.1093/mnras/stx3040}, 475, 624

\bibitem[\protect\citeauthoryear{{Nelson} et~al.,}{{Nelson}
  et~al.}{2019}]{tng-publicdata}
{Nelson} D.,  et~al., 2019, \mn@doi [Computational Astrophysics and Cosmology]
  {10.1186/s40668-019-0028-x}, \href
  {https://ui.adsabs.harvard.edu/abs/2019ComAC...6....2N} {6, 2}

\bibitem[\protect\citeauthoryear{{Okumura}, {Jing}  \& {Li}}{{Okumura}
  et~al.}{2009}]{Okumura2009}
{Okumura} T.,  {Jing} Y.~P.,   {Li} C.,  2009, \mn@doi [\apj]
  {10.1088/0004-637X/694/1/214}, \href
  {https://ui.adsabs.harvard.edu/abs/2009ApJ...694..214O} {694, 214}

\bibitem[\protect\citeauthoryear{{Okumura}, {Taruya}  \&
  {Nishimichi}}{{Okumura} et~al.}{2020}]{Okumura2020}
{Okumura} T.,  {Taruya} A.,   {Nishimichi} T.,  2020, \mn@doi [\mnras]
  {10.1093/mnras/staa718}, \href
  {https://ui.adsabs.harvard.edu/abs/2020MNRAS.494..694O} {494, 694}

\bibitem[\protect\citeauthoryear{{Padmanabhan} et~al.,}{{Padmanabhan}
  et~al.}{2008}]{2008ApJ...674.1217P}
{Padmanabhan} N.,  et~al., 2008, \mn@doi [\apj] {10.1086/524677}, \href
  {http://adsabs.harvard.edu/abs/2008ApJ...674.1217P} {674, 1217}

\bibitem[\protect\citeauthoryear{{Pier}, {Munn}, {Hindsley}, {Hennessy},
  {Kent}, {Lupton}  \& {Ivezi{\'c}}}{{Pier} et~al.}{2003}]{2003AJ....125.1559P}
{Pier} J.~R.,  {Munn} J.~A.,  {Hindsley} R.~B.,  {Hennessy} G.~S.,  {Kent}
  S.~M.,  {Lupton} R.~H.,   {Ivezi{\'c}} {\v Z}.,  2003, \aj, \href
  {http://adsabs.harvard.edu/cgi-bin/nph-bib_query?bibcode=2003AJ....125.1559P&amp;db_key=AST}
  {125, 1559}

\bibitem[\protect\citeauthoryear{{Pillepich} et~al.,}{{Pillepich}
  et~al.}{2018a}]{tng-methods}
{Pillepich} A.,  et~al., 2018a, \mn@doi [\mnras] {10.1093/mnras/stx2656}, \href
  {https://ui.adsabs.harvard.edu/abs/2018MNRAS.473.4077P} {473, 4077}

\bibitem[\protect\citeauthoryear{{Pillepich} et~al.,}{{Pillepich}
  et~al.}{2018b}]{pillepich2018illustristng}
{Pillepich} A.,  et~al., 2018b, \mn@doi [\mnras] {10.1093/mnras/stx3112}, \href
  {https://ui.adsabs.harvard.edu/abs/2018MNRAS.475..648P} {475, 648}

\bibitem[\protect\citeauthoryear{{Planck Collaboration} et~al.,}{{Planck
  Collaboration} et~al.}{2016}]{planck2016}
{Planck Collaboration} et~al., 2016, \mn@doi [\aap]
  {10.1051/0004-6361/201525830}, \href
  {https://ui.adsabs.harvard.edu/abs/2016A&A...594A..13P} {594, A13}

\bibitem[\protect\citeauthoryear{{Reid} et~al.,}{{Reid}
  et~al.}{2016}]{2016MNRAS.455.1553R}
{Reid} B.,  et~al., 2016, \mn@doi [\mnras] {10.1093/mnras/stv2382}, \href
  {https://ui.adsabs.harvard.edu/abs/2016MNRAS.455.1553R} {455, 1553}

\bibitem[\protect\citeauthoryear{{Reyes}, {Mandelbaum}, {Gunn}, {Nakajima},
  {Seljak}  \& {Hirata}}{{Reyes} et~al.}{2012}]{Reyes2012}
{Reyes} R.,  {Mandelbaum} R.,  {Gunn} J.~E.,  {Nakajima} R.,  {Seljak} U.,
  {Hirata} C.~M.,  2012, \mn@doi [\mnras] {10.1111/j.1365-2966.2012.21472.x},
  \href {http://adsabs.harvard.edu/abs/2012MNRAS.425.2610R} {425, 2610}

\bibitem[\protect\citeauthoryear{{Richards} et~al.,}{{Richards}
  et~al.}{2002}]{2002AJ....123.2945R}
{Richards} G.~T.,  et~al., 2002, \mn@doi [\aj] {10.1086/340187}, \href
  {http://adsabs.harvard.edu/abs/2002AJ....123.2945R} {123, 2945}

\bibitem[\protect\citeauthoryear{{Samuroff} et~al.,}{{Samuroff}
  et~al.}{2022}]{Samuroff2022}
{Samuroff} S.,  et~al., 2022, preprint, \href
  {https://ui.adsabs.harvard.edu/abs/2022arXiv221211319S} {} (\mn@eprint
  {arXiv} {2212.11319})

\bibitem[\protect\citeauthoryear{{S{\'a}nchez} et~al.,}{{S{\'a}nchez}
  et~al.}{2017}]{Sanchez2017}
{S{\'a}nchez} A.~G.,  et~al., 2017, \mn@doi [\mnras] {10.1093/mnras/stw2443},
  \href {https://ui.adsabs.harvard.edu/abs/2017MNRAS.464.1640S} {464, 1640}

\bibitem[\protect\citeauthoryear{{Secco} et~al.,}{{Secco}
  et~al.}{2022}]{Secco2022}
{Secco} L.~F.,  et~al., 2022, \mn@doi [\prd] {10.1103/PhysRevD.105.023515},
  \href {https://ui.adsabs.harvard.edu/abs/2022PhRvD.105b3515S} {105, 023515}

\bibitem[\protect\citeauthoryear{{Singh} \& {Mandelbaum}}{{Singh} \&
  {Mandelbaum}}{2016}]{Singh2016ia}
{Singh} S.,  {Mandelbaum} R.,  2016, \mn@doi [\mnras] {10.1093/mnras/stw144},
  \href {https://ui.adsabs.harvard.edu/abs/2016MNRAS.457.2301S} {457, 2301}

\bibitem[\protect\citeauthoryear{{Singh}, {Mandelbaum}  \& {More}}{{Singh}
  et~al.}{2015}]{Singh2015}
{Singh} S.,  {Mandelbaum} R.,   {More} S.,  2015, \mn@doi [\mnras]
  {10.1093/mnras/stv778}, \href
  {http://adsabs.harvard.edu/abs/2015MNRAS.450.2195S} {450, 2195}

\bibitem[\protect\citeauthoryear{{Singh}, {Mandelbaum}  \&
  {Brownstein}}{{Singh} et~al.}{2017a}]{2017MNRAS.464.2120S}
{Singh} S.,  {Mandelbaum} R.,   {Brownstein} J.~R.,  2017a, \mn@doi [\mnras]
  {10.1093/mnras/stw2482}, \href
  {https://ui.adsabs.harvard.edu/abs/2017MNRAS.464.2120S} {464, 2120}

\bibitem[\protect\citeauthoryear{{Singh}, {Mandelbaum}, {Seljak}, {Slosar}  \&
  {Vazquez Gonzalez}}{{Singh} et~al.}{2017b}]{Singh2017cov}
{Singh} S.,  {Mandelbaum} R.,  {Seljak} U.,  {Slosar} A.,   {Vazquez Gonzalez}
  J.,  2017b, \mn@doi [\mnras] {10.1093/mnras/stx1828}, \href
  {http://adsabs.harvard.edu/abs/2017MNRAS.471.3827S} {471, 3827}

\bibitem[\protect\citeauthoryear{{Singh}, {Alam}, {Mandelbaum}, {Seljak},
  {Rodriguez-Torres}  \& {Ho}}{{Singh} et~al.}{2019}]{2019MNRAS.482..785S}
{Singh} S.,  {Alam} S.,  {Mandelbaum} R.,  {Seljak} U.,  {Rodriguez-Torres} S.,
    {Ho} S.,  2019, \mn@doi [\mnras] {10.1093/mnras/sty2681}, \href
  {https://ui.adsabs.harvard.edu/abs/2019MNRAS.482..785S} {482, 785}

\bibitem[\protect\citeauthoryear{{Singh}, {Yu}  \& {Seljak}}{{Singh}
  et~al.}{2021}]{Singh2021}
{Singh} S.,  {Yu} B.,   {Seljak} U.,  2021, \mn@doi [\mnras]
  {10.1093/mnras/staa3263}, \href
  {https://ui.adsabs.harvard.edu/abs/2021MNRAS.501.4167S} {501, 4167}

\bibitem[\protect\citeauthoryear{{Smee} et~al.,}{{Smee}
  et~al.}{2013}]{Smee:2013}
{Smee} S.~A.,  et~al., 2013, \mn@doi [\aj] {10.1088/0004-6256/146/2/32}, \href
  {http://adsabs.harvard.edu/abs/2013AJ....146...32S} {146, 32}

\bibitem[\protect\citeauthoryear{{Smith} et~al.,}{{Smith}
  et~al.}{2002}]{2002AJ....123.2121S}
{Smith} J.~A.,  et~al., 2002, \aj, \href
  {http://adsabs.harvard.edu/cgi-bin/nph-bib_query?bibcode=2002AJ....123.2121S&amp;db_key=AST}
  {123, 2121}

\bibitem[\protect\citeauthoryear{{Smith} et~al.,}{{Smith}
  et~al.}{2003}]{Smith2003}
{Smith} R.~E.,  et~al., 2003, \mn@doi [\mnras]
  {10.1046/j.1365-8711.2003.06503.x}, \href
  {http://adsabs.harvard.edu/abs/2003MNRAS.341.1311S} {341, 1311}

\bibitem[\protect\citeauthoryear{{Springel}}{{Springel}}{2010}]{arepo}
{Springel} V.,  2010, \mn@doi [\mnras] {10.1111/j.1365-2966.2009.15715.x},
  \href {https://ui.adsabs.harvard.edu/abs/2010MNRAS.401..791S} {401, 791}

\bibitem[\protect\citeauthoryear{{Springel}, {White}, {Tormen}  \&
  {Kauffmann}}{{Springel} et~al.}{2001}]{subfind}
{Springel} V.,  {White} S. D.~M.,  {Tormen} G.,   {Kauffmann} G.,  2001,
  \mn@doi [\mnras] {10.1046/j.1365-8711.2001.04912.x}, \href
  {https://ui.adsabs.harvard.edu/abs/2001MNRAS.328..726S} {328, 726}

\bibitem[\protect\citeauthoryear{Springel et~al.}{Springel
  et~al.}{2018}]{Springel2017illustristng}
Springel V.,  et~al., 2018, \mn@doi [Mon. Not. Roy. Astron. Soc.]
  {10.1093/mnras/stx3304}, 475, 676

\bibitem[\protect\citeauthoryear{{Strauss} et~al.,}{{Strauss}
  et~al.}{2002}]{2002AJ....124.1810S}
{Strauss} M.~A.,  et~al., 2002, \aj, \href
  {http://adsabs.harvard.edu/cgi-bin/nph-bib_query?bibcode=2002AJ....124.1810S&db_key=AST}
  {124, 1810}

\bibitem[\protect\citeauthoryear{{Takahashi}, {Sato}, {Nishimichi}, {Taruya}
  \& {Oguri}}{{Takahashi} et~al.}{2012}]{Takahashi2012}
{Takahashi} R.,  {Sato} M.,  {Nishimichi} T.,  {Taruya} A.,   {Oguri} M.,
  2012, \mn@doi [\apj] {10.1088/0004-637X/761/2/152}, \href
  {http://adsabs.harvard.edu/abs/2012ApJ...761..152T} {761, 152}

\bibitem[\protect\citeauthoryear{{Troxel} \& {Ishak}}{{Troxel} \&
  {Ishak}}{2015}]{Troxel2015}
{Troxel} M.~A.,  {Ishak} M.,  2015, \mn@doi [\physrep]
  {10.1016/j.physrep.2014.11.001}, \href
  {http://adsabs.harvard.edu/abs/2015PhR...558....1T} {558, 1}

\bibitem[\protect\citeauthoryear{{Tucker} et~al.,}{{Tucker}
  et~al.}{2006}]{2006AN....327..821T}
{Tucker} D.~L.,  et~al., 2006, \mn@doi [Astronomische Nachrichten]
  {10.1002/asna.200610655}, \href
  {http://adsabs.harvard.edu/abs/2006AN....327..821T} {327, 821}

\bibitem[\protect\citeauthoryear{{Vlah} \& {White}}{{Vlah} \&
  {White}}{2019}]{Vlah2019}
{Vlah} Z.,  {White} M.,  2019, \mn@doi [\jcap] {10.1088/1475-7516/2019/03/007},
  \href {https://ui.adsabs.harvard.edu/abs/2019JCAP...03..007V} {2019, 007}

\bibitem[\protect\citeauthoryear{{Vlah}, {Chisari}  \& {Schmidt}}{{Vlah}
  et~al.}{2020}]{2020JCAP...01..025V}
{Vlah} Z.,  {Chisari} N.~E.,   {Schmidt} F.,  2020, \mn@doi [\jcap]
  {10.1088/1475-7516/2020/01/025}, \href
  {https://ui.adsabs.harvard.edu/abs/2020JCAP...01..025V} {2020, 025}

\bibitem[\protect\citeauthoryear{{Weinberg}, {Mortonson}, {Eisenstein},
  {Hirata}, {Riess}  \& {Rozo}}{{Weinberg} et~al.}{2013}]{Weinberg2013}
{Weinberg} D.~H.,  {Mortonson} M.~J.,  {Eisenstein} D.~J.,  {Hirata} C.,
  {Riess} A.~G.,   {Rozo} E.,  2013, \mn@doi [\physrep]
  {10.1016/j.physrep.2013.05.001}, \href
  {http://adsabs.harvard.edu/abs/2013PhR...530...87W} {530, 87}

\bibitem[\protect\citeauthoryear{{Weinberger} et~al.,}{{Weinberger}
  et~al.}{2017}]{tng-agn}
{Weinberger} R.,  et~al., 2017, \mn@doi [\mnras] {10.1093/mnras/stw2944}, \href
  {https://ui.adsabs.harvard.edu/abs/2017MNRAS.465.3291W} {465, 3291}

\bibitem[\protect\citeauthoryear{{Yao}, {Ishak}, {Troxel}  \& {LSST Dark Energy
  Science Collaboration}}{{Yao} et~al.}{2019}]{Yao2019}
{Yao} J.,  {Ishak} M.,  {Troxel} M.~A.,   {LSST Dark Energy Science
  Collaboration} 2019, \mn@doi [\mnras] {10.1093/mnras/sty3188}, \href
  {https://ui.adsabs.harvard.edu/abs/2019MNRAS.483..276Y} {483, 276}

\bibitem[\protect\citeauthoryear{{York} et~al.,}{{York}
  et~al.}{2000}]{2000AJ....120.1579Y}
{York} D.~G.,  et~al., 2000, \aj, \href
  {http://adsabs.harvard.edu/cgi-bin/nph-bib_query?bibcode=2000AJ....120.1579Y&db_key=AST}
  {120, 1579}

\bibitem[\protect\citeauthoryear{{van den Busch} et~al.,}{{van den Busch}
  et~al.}{2022}]{vandenBusch2022}
{van den Busch} J.~L.,  et~al., 2022, \mn@doi [\aap]
  {10.1051/0004-6361/202142083}, \href
  {https://ui.adsabs.harvard.edu/abs/2022A&A...664A.170V} {664, A170}

\makeatother
\end{thebibliography}
	
\end{document}